\documentclass{SciPost}

\hypersetup{
    colorlinks,
    linkcolor={red!50!black},
    citecolor={blue!50!black},
    urlcolor={blue!80!black}
}

\usepackage[bitstream-charter]{mathdesign}
\DeclareSymbolFont{usualmathcal}{OMS}{cmsy}{m}{n}
\DeclareSymbolFontAlphabet{\mathcal}{usualmathcal}

\usepackage[english]{babel}
\usepackage{amsmath}
\usepackage{color}
\usepackage{xcolor}
\usepackage{multirow}
\usepackage[normalem]{ulem}
\usepackage{booktabs}
\usepackage{enumitem}
\usepackage{etoolbox}

\newcommand{\CG}[1]{\textcolor{green}{\bf{#1}}}

\newcommand{\nog}[1]{\textcolor{green}{\sout{#1}}}

\newcommand{{\au}}[0]{{$a_0$}}

\fancypagestyle{SPstyle}{
  \fancyhf{}
  
  \fancyfoot[C]{\textbf{\thepage}}
}

\fancypagestyle{plain}{
  \fancyhf{}

  \fancyfoot[C]{\textbf{\thepage}}
}

\makeatletter
\patchcmd{\@maketitle}{\includegraphics[width=25mm]{SciPost_logo.pdf}}{}{}{}
\makeatother

\makeatletter
\fancypagestyle{SPstyle}{
  \fancyhf{}
  
  \fancyfoot[C]{\textbf{\thepage}}
}
\fancypagestyle{plain}{
  \fancyhf{}

  \fancyfoot[C]{\textbf{\thepage}}
}

\renewcommand{\@maketitle}{%
  \newpage
  \null
  \vskip 1em%
  \begin{center}%
    {\LARGE \bfseries \@title \par}%
    \vskip 1.5em%
    {\large
      \lineskip .5em%
      \begin{tabular}[t]{c}%
        \@author
      \end{tabular}\par}%
  \end{center}%
  \par
  \vskip 1.5em%
}
\makeatother

\makeatletter
\patchcmd{\@maketitle}{\includegraphics[width=25mm]{SciPost_logo.pdf}}{}{}{}
\makeatother
\begin{document}

\pagestyle{SPstyle}

\begin{center}{\Large \textbf{\color{scipostdeepblue}{
Inelastic confinement-induced resonances in multiwell potentials
}}}\end{center}

\begin{center}\textbf{
Fabio Revuelta\textsuperscript{1$\star$} and
Alejandro Saenz\textsuperscript{2 $\dag$}
}\end{center}

\begin{center}
{\bf 1} Grupo de Sistemas Complejos, 
%Escuela T\'ecnica Superior de
E.\,T.\,S. Ingenier\'ia Agron\'omica, Alimentaria y de Biosistemas, Universidad Polit\'ecnica de Madrid, Avenida Puerta de Hierro 2, 28040 Madrid, Spain.
\\
{\bf 2} AG Moderne Optik, Institut f\"ur Physik, Humboldt-Universit\"at zu Berlin, Newtonstrasse 15, 12489 Berlin, Germany.
\\[\baselineskip]
$\star$ \href{mailto:fabio.revuelta@upm.es}{\small fabio.revuelta@upm.es,}
$\dag$ \href{mailto:alejandro.saenz@physik.hu-berlin.de}{\small alejandro.saenz@physik.hu-berlin.de}
\end{center}

\section*{\color{scipostdeepblue}{Abstract}}
\textbf{\boldmath{%
Recently, resonance features associated with multiwell effects have been reported in a fascinating experiment conducted with an ultracold-atom quantum simulator
[Y.\,K.~Lee, 
H.~Lin, and W.~Ketterle, \emph{Phys. Rev. Lett.} \textbf{131}, 213001 (2023)].
%~\cite{cold:lee23}.
The results were explained as a consequence of superexchange, which determines the spin dynamics in a singly-occupied
Mott insulator, by allowing for the virtual population of two particles on the same site. In this work, 
%\nog{we provide}
a complementary interpretation of the resonance features is provided based on
%\nog{as a consequence of}
inelastic confinement-induced resonances, manifested as avoided crossings in the two-particle spectrum, showing 
%\no{excellent} 
very good
agreement with the experimental results. The comparison of the results of the single- and double-well descriptions of the optical trapping potential confirms the multiwell origin of the experimental features. 
However, the double-well model is already shown to be sufficient for providing an explanation of the apparent symmetry violation that 
%\nog{is observed}
occurs within 
%\nog{the} 
single-well
%\nog{approach} 
models.
Furthermore, certain limitations of the harmonic approximation are discussed.
}}

\vspace{\baselineskip}

\noindent\textcolor{white!90!black}{%
\fbox{\parbox{0.975\linewidth}{%
\textcolor{white!40!black}{\begin{tabular}{lr}%
  \begin{minipage}{0.6\textwidth}%
    {\small Copyright attribution to authors. \newline
    This work is a submission to SciPost Physics. \newline
    License information to appear upon publication. \newline
    Publication information to appear upon publication.}
  \end{minipage} & \begin{minipage}{0.4\textwidth}
    {\small Received Date \newline Accepted Date \newline Published Date}%
  \end{minipage}
\end{tabular}}
}}
}

%\linenumbers

\vspace{10pt}
\noindent\rule{\textwidth}{1pt}
\tableofcontents
\noindent\rule{\textwidth}{1pt}
\vspace{10pt}

% =========================================================================
% ARTICLE CONTENTS
% =========================================================================

%==================================================================
% INTRODUCTION
\section{Introduction} \label{sec:intro}
%==================================================================
The study of ultracold atomic gases
has noticeably extended over the last years~\cite{cold:peth02,
cold:lewe12}.
%cold:vule00, cold:peth02, cold:pita03, cold:lewe12
%becoming a cornerstone for modern quantum simulation 
%\cite{cold:nielsen10, cold:gross17, cold:hof18, cold:arg19, cold:scha20, cold:arg25}
The reason is twofold,
and relies on the excellent degree of control that can be achieved
over the interactions and the confinement~\cite{cold:veyr24}.

%FR
On the one hand, 
the atom-atom interaction can be tuned almost at will
thanks to the magnetic Feshbach resonances (FRs)~\cite{cold:wall12, cold:juli04, cold:mies00, cold:schu05}.
These resonances occur when 
the energy of a molecular bound state
is shifted under an external magnetic field
thanks to the hyperfine structure of the atoms,
becoming \emph{resonant} with
the energy of two colliding atoms in free space~\cite{cold:simo09, cold:chin10}.
A similar coupling can be also optically induced
\cite{cold:fedi96, cold:theis04, cold:thal05, cold:sand17}.
Remarkably, within the ultracold regime, s-wave scattering dominates 
(except for indistinguishable Fermions).
Thus the atom-atom interaction can be characterized by a single
parameter, the s-wave scattering length $a_{\rm sc}$~\cite{cold:bold02, cold:peth02},
which diverges at FRs. As a consequence,
the effective interaction between two particles
%\nog{changes} 
can be tuned from strongly attractive,
if 
$a_{\rm sc}$ presents large
negative
values,
%$a_{\rm sc} \rightarrow \infty $
%($a_{\rm sc} << 0 $)
to repulsive, if $a_{\rm sc}$ has large 
positive
values,
%$a_{\rm sc} \rightarrow -\infty $, 
or {\it vice versa}.

% OL, OT...
%Similarly,
On the other hand,
the confinement of the atoms has also added an exquisite degree of control~\cite{cold:taka05}.
Atoms can be distributed at arbitrary positions 
in space with optical lattices
created with counter propagating laser fields~\cite{cold:jaks98, cold:bloc05, cold:lewe07, cold:meac98}.
Furthermore, 
when the laser intensities in one or
two dimensions are substantially lowered,
a three-dimensional (3D) lattice~\cite{cold:koeh05, cold:ospe06a, cold:schn08, cold:cape23}
 effectively reduces to a
quasi-one 
%(Q1D)
dimensional (Q1D) elongated trap~\cite{cold:cris02, cold:gran04, cold:kino05a, cold:klau01, cold:mora05, cold:rigo05b, cold:naid06, cold:lee23}
or to a quasi-two 
%(Q2D)
dimensional (Q2D) planar trap~\cite{cold:froh11, cold:wout03, cold:naid06},
where the motion of the atoms is effectively restricted to take place
%effectively takes place
%dramatically facilitated
over tubes or planes,
respectively.
Indeed,
optical lattices have played a central role in the creation
of quantum-gas microscopes~\cite{cold:weit11, cold:scha20, cold:veyr24},
where the positions of single atoms are accurately determined and monitored.
These extraordinary tools are also complemented with optical
tweezers, also known
as  microtraps~\cite{cold:serw11,cold:kauf12},
which are very tiny traps,
in which few atoms are strongly confined.
The combination of several optical tweezers
has also permitted new methods to synthesize ultracold molecules~\cite{cold:sala13, cold:yicha21,
cold:rutt23, cold:gutt25}
with unforeseen precision.
Likewise,
the so-called ~tweezer arrays~\cite{cold:barr18}
have opened new routes to study
many-particle correlated matter,
such as quantum many-body scars~\cite{cold:turn18}.
%cold:serb21
Moreover,
the number of atoms can be also arbitrarily controlled
in experiments all over the world,
ranging from solely a few particles~\cite{cold:sala13, cold:wenz13, cold:bayha20, cold:rutt23}
to large atomic samples
constituted by thousands 
%or even millions
of atoms~\cite{cold:hall10b, cold:cape23}.
Likewise,
different atomic species~\cite{cold:sowin19, cold:rutt23, cold:gutt25}
or atoms with different internal states~\cite{cold:sala13, cold:lee23}
can be used with high flexibility.
In summary,
the high degree of control that is nowadays
achieved in ultracold atoms has transformed them into
one of the most promising platforms
%this platform as one of the most promising ones
for quantum-simulation
 and even quantum-computation
applications
%NEW (NOV'26)
\cite{cold:jaks98, cold:grei02, cold:schn12, cold:ber17, cold:bluv26}.
%ORIG
%\cite{cold:nielsen10, cold:gross17, %cold:hof18, cold:arg19, cold:scha20, %cold:arg25}
%
The combination of tunable interactions and
external optical confinements is, nevertheless,
far from trivial as it %~\cite{cold:petr01, cold:naid06, cold:idzi06, cold:lian08, cold:sowin19, cold:peng20},
changes the system behavior dramatically~\cite{cold:petr01, cold:naid06, cold:idzi06, cold:lian08, cold:sowin19, cold:peng20,cold:petr00, cold:petr00b, cold:petr01, cold:fedi04, cold:koeh05, cold:stoe06, cold:dien06b, cold:hall10b, cold:kest10, cold:kest10b, cold:stec11, cold:sala12, cold:wall12}.

In a seminal work, the scattering process 
under external confinement was analyzed by Olshanii for 
two interacting particles 
in a harmonic Q1D wave guide~\cite{cold:olsh98}, resulting in the 
introduction of an effective one-dimensional interaction parameter 
that is a unique function of the s-wave scattering length.
%\nog{Later on}
In a subsequent work,
Bergeman \emph{et al.}~\cite{cold:berg03} %\nog{introduced}
discussed the so-called 
(elastic) confinement-induced resonances
(CIRs)
which occur if the ratio of the scattering length 
and the confinement strength
acquires a specific value for which the effective 
1D interaction parameter diverges. 
Similarly, also under 
Q2D
%quasi two-dimensional 
confinement (elastic) CIRs
were predicted \cite{cold:petr00} and 
thereafter experimentally observed~\cite{cold:froh11}.
However, the theory of elastic CIRs
could not explain the Q1D results reported by Haller \emph{et al.}\ in 
a groundbreaking experiment where the 
resonant loss features of an ultracold atomic sample 
split under anisotropic Q1D confinement.
Nowadays, the results of the experiment
in Ref.~\cite{cold:hall10b} 
are recognized to be a consequence of 
a different type of resonances~\cite{cold:sala12}:
the {\it inelastic} confinement-induced resonances (ICIRs).
ICIRs are ubiquitous in traps as they are caused by the coupling between the relative-motion (RM) and the center-of-mass (CM)
%\nog{coordinates} 
degrees of freedom
by, e.\,g., possible anharmonicities of the external confinement  
\cite{cold:bold05,cold:pean05,cold:gris07,cold:sala12}, 
which explains
why 
%in this case 
they were named anharmonicity-induced resonances 
by Kestner and Duan~\cite{cold:kest10, cold:kest10b}.

Notably, the CM-RM coupling can also arise under harmonic
confinement~\cite{cold:gris09}
if different atomic species are considered~\cite{cold:baro24},
or if 
the same species in different hyperfine states 
is trapped with different frequencies. Moreover,
two‑particle systems under strictly harmonic confinement can 
also exhibit trap‑induced 
resonances~\cite{cold:stoc03, cold:kryc09, cold:okeyo26inwork}
if the
traps are not equally centered in space.
Also in this type of ICIRs, 
%\no{these}
resonances appear as avoided crossings in
the two‑particle spectrum when a molecular bound state becomes
resonant with a trap eigenstate,
though in this case the resonance is controlled by
the relative separation of the trapping potentials.

The existence of ICIRs is nowadays accepted 
within the ultracold community,
as they have been observed in systems of different shape
and dimension~\cite{cold:hall10b, cold:cape23}, 
including different atomic species~\cite{cold:lampo10}. 
In fact, ICIRs have 
also enabled the coherent molecule formation 
of two Fermionic atoms~\cite{cold:sala13}, a process that cannot 
be induced by elastic CIRs.
Furthermore, their occurrence in solid-state devices has been also predicted~\cite{cold:trop15}. 
In this case, the interaction is not 
of van-der-Waals kind but Coulombic.
Their existence under 3D confinement, as recently 
demonstrated by Capecchi \emph{et al.}~\cite{cold:cape23},
clearly distinguishes  
them from elastic CIRs that can only occur in 
Q1D or Q2D systems.

In this work,
we revisit the experiment recently performed by Lee \emph{et al.}~\cite{cold:lee23}.
By considering a simple theoretical model comprising two ultracold atoms
that interact via a realistic
\emph{ab-initio}  potential,
we provide an alternative explanation of 
the experimental results that does not rely on a solid-state inspired
interpretation but on a purely few-body molecular perspective.
The agreement between the loss features detected in the experiment, 
which were associated in Ref.~\cite{cold:lee23}
with lattice-induced resonances~\cite{cold:stec11},
and the predictions rendered by our numerical simulations 
is very good.
Remarkably, by considering a double-well (DW) potential the apparent symmetry violation that is discussed in the context of the adopted
harmonic model in the Ref.~\cite{cold:lee23} is also lifted.
In addition, we provide a simple and transparent explanation 
for the experimental observation of ICIRs under Q1D confinement 
and its non-detection under isotropic 3D confinement.
Noticeably, the reasonable performance of the single-well (SW) 
description of the system is also discussed,
and its accuracy is detailed in comparison with the DW results.
Both, harmonic and anharmonic potentials 
%Both harmonic (SW2) and anharmonic potentials of order~6 (SW6)
are considered within the SW descriptions of the optical trap.
In this way, we show that the so called lattice-induced resonances
that are discussed as a consequence of tunneling between neighboring sites
have, indeed, the same origin as the ICIRs~\cite{cold:sala12}: 
the RM-CM coupling. 

The outline of this work is as follows. First,
Sec.~\ref{sec:system_methods} 
presents the studied physical system,
outlining the key features of the experimental setup,
the theoretical models employed,
and the numerical methods implemented.
Second, 
Sec.~\ref{sec:results} contains the results of our study
along with the corresponding discussion.
Finally,
we conclude this paper 
in Sec.~\ref{sec:summary}
with a brief summary and outlook.

%==================================================================
% SYSTEM AND METHODS
\section{System and methods} \label{sec:system_methods}
%==================================================================

This section
%\nog{is devoted to} 
describes the system and the methods considered
for its theoretical description.
%\nog{First, }
Section~\ref{sec:experiment}
briefly summarizes the main characteristics
and the current interpretation of
%performed by Y.~K. Lee, H. Lin and W. Ketterle 
%detailed in
the experiment reported in the Ref.~\cite{cold:lee23}.
%\nog{Second, }
Section~\ref{sec:models} is devoted to 
the two-particle models developed for this work.
%to analyze the experiment.
%that have been developed to adequately describe
%the experimental outcomes within the ICIR framework.
%
%\CR{\emph{[FR: remove the next sentence as it is detailed in Sec. II.A?]}
%by considering
%(i) a realistic \emph{ab-initio} interaction,
%and
%(ii) an optical lattice solely determined by the experimental parameters
%of the confining lasers.}
%(atomic polarizabilities and laser parameters).
%where multiwell effects on lattice-induced resonances is reported.
%
%\nog{Third,}
Finally, 
Sec.~\ref{sec:numerics} briefly summarizes the
%model considered to unveil
%the resonance features that are experimentally observed
%using an alternative perspective which is based on a double-well (DW)
numerical scheme that has been used.
%considered to 
%identify the ICIRs as avoided crossings in the two-particle spectrum.
%
%As discussed below,
%one of the main advantages of this approach
%is that
%the avoided crossings in the energy spectrum for
%a two-particle system straightforward reproduce 
%in a rather accurate way
%the positions of the resonance features that are experimentally
%observed with no further tuning or fitting parameters.

%\newpage . \newpage
%==================================================================
\subsection{Summary of the experiment %\nog{conducted}
in the Ref.~\cite{cold:lee23}}
\label{sec:experiment}
%==================================================================

\begin{figure}[hbtp!]
\includegraphics[width=0.95\columnwidth]{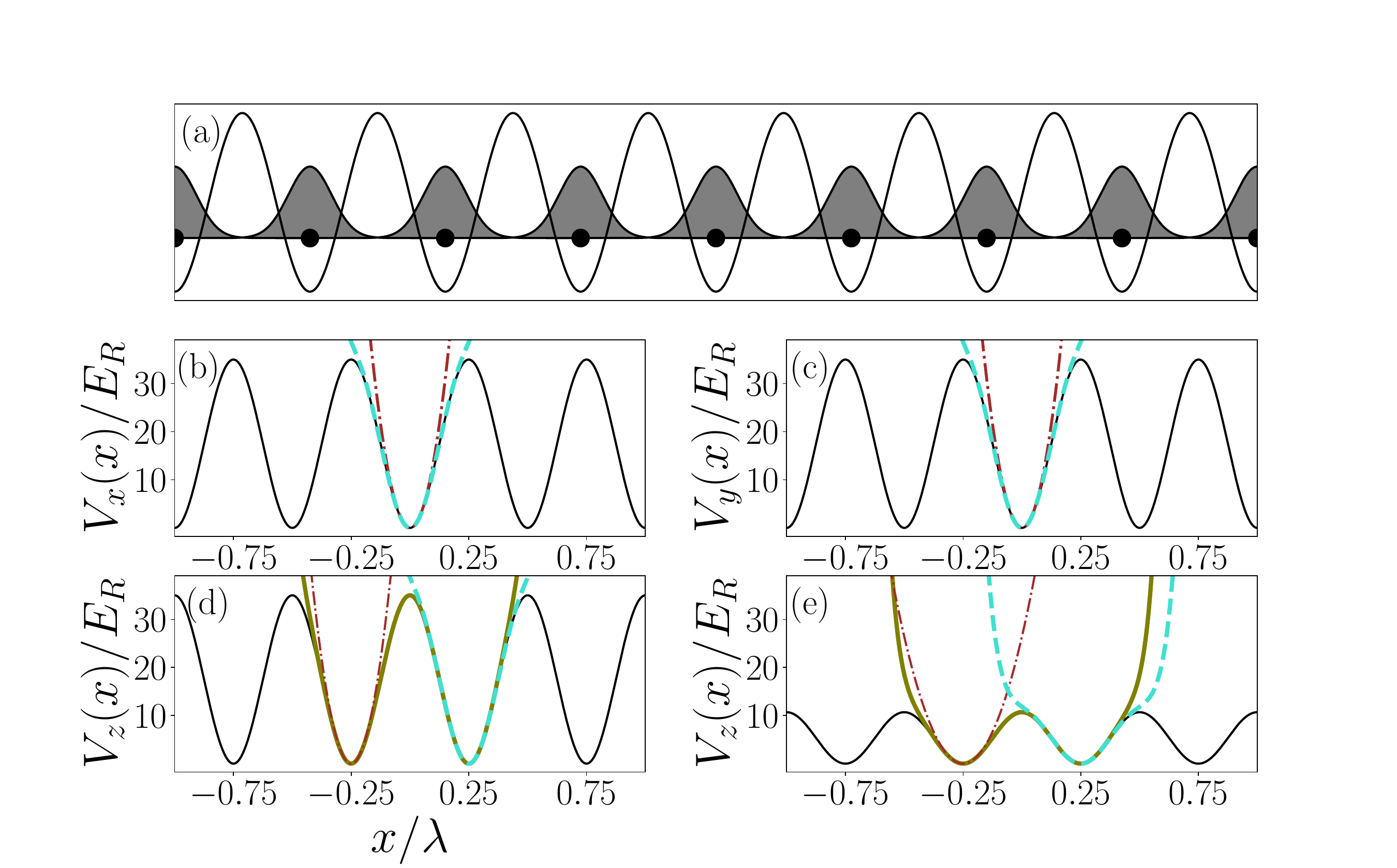}
\caption{Ultracold atoms trapped in an optical lattice
and the considered models. 
(a) Sketch of a Mott insulator state with a single-site occupancy
trapped in an optical lattice
[cf. Eqs.~\eqref{eq:Vsin2} and~\eqref{eq:Vcos2}]. 
(b) Single-well potential along the~$x$ direction for
a \emph{large} trap depth of~$V_x = 35~E_R$,
comparing the harmonic approximation of Eq.~\eqref{eq:Vharm} (dashed brown)
with the sextic potential of Eq.~\eqref{eq:Vsw} (dash-dotted cyan).
(c) Same as~(b) for the~$y$ direction.
(d) Same as~(b) along the~$z$ direction,
along with the double-well description 
given by Eq.~\eqref{eq:Vdw} (solid olive).
(e) Same as~(d) for
a \emph{small} trap depth 
of~$V_z = 10.7~E_R$.}
%}
\label{fig:pot}
\end{figure}

In the experiment 
%\no{discussed}
presented
in the Ref.~\cite{cold:lee23},
a Mott insulator~\cite{cold:syas07, cold:robe08} 
of $^7$Li
%lithium-7
ultracold atoms
trapped in an optical lattice
with single-site occupancy is first created,
as sketched in Fig.~\ref{fig:pot}(a).
The atoms are in the two lowest hyperfine states
(labeled as $a$ and $b$),
%and, as a consequence,
%the system resembles
mimicking a two-component Bose-Hubbard model~\cite{cold:lewe12}.
%
%\nog{Then} 
Under these conditions, 
three-body processes are very rare 
because of
single-site occupancy and the short-range behavior of the interparticle 
van-der-Waals interaction.

%
%--------------------------------------------------------------------------------------------
% Figure: POTENTIALS
%--------------------------------------------------------------------------------------------
%--------------------------------------------------------------------------------------

%\newpage
%\nog{First}
In the first experimental setting,
the atoms are confined in an
%a strictly 
isotropic 3D optical lattice
%trap
with potential depths
equal to~$V_x = V_y = V_z = 35 \, E_R$,
with~$E_R=h^2/(2 m \lambda^2) \approx 25$\,kHz
%ER= 25.172 kHz
the recoil energy,
and~$\lambda=1064$\,nm the wave length;
these parameters render
%which renders isotropic
trapping frequencies
equal to
$\omega_x = \omega_y = \omega_z = $
$ 2 \pi \times 297.2$\,kHz.
% TwoAtInOL:
%
% Frequency w_x in KHz        =    297.208938134159
% Frequency w_y in KHz        =    297.208938134159
% Frequency w_z in KHz        =    164.357873904878
%
%and~$\lambda=1064$\,nm the wave length;
%the corresponding trapping frequencies are related as
%$\omega_x = \omega_y = 1.8 \omega_z$.

%\nog{Second}
In the second setting, 
by lowering the laser intensity in the~$z$ direction,
the optical lattice becomes strongly elongated,
creating
%forming
%then becoming
a Q1D confinement
%optical lattice
%trap
with potential depths
equal to~$V_x = V_y = 35 E_R$, and~$V_z= 10.7 \, E_R$. In
%\nog{; in}\CG{. In}
%with~$E_R=h^2/(2 m \lambda^2) \approx 25$\,kHz
%the recoil energy,
%and~$\lambda=1064$\,nm the wave length;
this case,
the corresponding trapping frequencies are related as
$\omega_x = \omega_y =$
$ 1.8 \, \omega_z$.

%\newpage
The system symmetry
%This configuration
enables the realization of the spin-1/2 XXZ Heisenberg model~\cite{cold:lewe12},
where the effective spin-spin interactions are a combination
of superexchange processes and off-site interactions.
In order to characterize these
% spin-spin
interactions,
%
%which depends on spin-spin interactions due to superexchange and off-site contributions,
%$\Delta = (J_{bb} + J_0)/Jxy$,
the researchers 
%\nog{of}
in the Ref.~\cite{cold:lee23}
determined the spin-anisotropy parameter
%~$\Delta$, 
by exploiting the enhanced lifetime of spin helix states~\cite{cold:popkov21, cold:jepsen22}
at a specific wave vector.
The spin-anisotropy parameter
%$\Delta$ parameter
is influenced by interactions among all three spin-state combinations ($aa$, $ab$, and $bb$).
This parameter
%The spin anisotropy parameter
was experimentally measured
by tuning the $bb$ interaction around one narrow and
one wide magnetic FR~\cite{cold:wall12, cold:juli04, cold:mies00, cold:schu05}
associated with the $bb$ atoms,
which, 
remarkably,
does not affect 
the $aa$ and $ab$ spin-spin interactions.

Within the wide FR,
three dispersive  features are observed at
892.249, 893.182, and 895.537\,G.
An additional dispersive feature more difficult to probe
is found at 845.459\,G.
The corresponding scattering lengths are equal to
4632\,$a_0$, 2287\,$a_0$, $-2662\,a_0$, and $-2593\,a_0$~\cite{IntComm2022},
respectively,
$a_0$ being the Bohr radius.
% asc = -2593 a0 corresponds to the narrow resonance located at 845.459\,G.
% More precise:
% asc =  4631.93
% asc =  2287.42
% asc = -2593.13 
% asc = -2661.96
%
No dispersive feature is observed in the 3D experimental setup.
As reported in the Ref.~\cite{cold:lee23},
none of those results can be associated with already
known FRs. 
Conversely,
the comparison with the results predicted by a simplified model 
of two particles confined in an (isotropic or anisotropic) harmonic trap
and interacting through a contact potential,
the authors relate
them to lattice-induced resonances~\cite{cold:stec11}.
Despite using a harmonic model, %\nog{description of the real system},
the authors attribute the observed features to multiwell,
i.\,e., to 
lattice effects, by
searching for crossings between the molecular states 
formed under harmonic confinement
%that must be formed
and
the zero-energy point related to the initial Mott-insulator state.
%delocalized over the whole lattice potential.
%\nog{The key point lies on a}
Within this interpretation,
the key 
point
%\nob{process}
is a resonant 
directional tunneling of one atom from one well to the neighboring site
in order to create a bound molecule,
%\nog{in}
in a process that
%\nob{which}
breaks the parity symmetry 
of the system. From the 
%\nog{energetic} 
energy point of view,
the 
%\nog{interaction}
bond formation is possible because 
%the effective interaction between the initial Mott state and
in the final molecular states the (negative) binding energy is balanced 
by the (positive) CM energy of the created molecule,
which enables the crossing 
%\nog{with the zero-energy point associated}
with the initial Mott-insulator state of two atoms 
at neighbor sites. 
Although the importance of the 
CM-RM
coupling 
is emphasized, the authors in \cite{cold:lee23} conclude that the 
resonances cannot be ICIRs.
%\nog{Noticeably}
Notably, 
the authors in Ref.~\cite{cold:lee23} 
emphasize the simplicity of their harmonic-trap model, as in lattices 
many of the energy level crossings become avoided and the bands attain 
finite widths.

%==================================================================
\subsection{Theoretical models}
%\subsection{System and models}
%\subsection{Summary of the experiment conducted in the Ref.~\cite{cold:lee23}
\label{sec:models}
%==================================================================
We model the system of interest
%\nog{studied} 
using the two-particle Hamiltonian
\begin{equation}
\label{eq:H}
   \hat{\mathcal{H}} \left( \mathbf{r}_1,  \mathbf{r}_2 \right)
= \sum_{i=1}^2 \left[ - \frac{\hbar^2}{2m} \mathbf{\hat{p}}_i^2 +
\hat{\mathcal{V}}_{\rm opt} \left( \mathbf{r}_i \right) \right] +
%\hat{\mathcal{V}}_{\rm opt}^{(i)} \left( \mathbf{r}_i \right) \right] +
\hat{U} \left( \mathbf{r}_1,  \mathbf{r}_2 \right),
%{   {H}}\,(\, \mathbf{r}_1\,, \mathbf{r}_2 \,)\, = 
\end{equation}
where the sum accounts for the 
single-particle terms related 
to the kinetic-energy operator
and to the potential-energy operator exerted by the optical lattice
for each of the particles,
%the first one to the kinetic energy,
%and the second one to the potential energy exerted by the optical lattice
%for each of the particles,
% trapping potential
while the 
last term is associated with the interparticle 
interaction.
In the present study,
this interaction is mostly described using a realistic
\emph{ab-initio} 
Born-Oppenheimer curve for a Li-Li dimer. As detailed below,
although the interatomic potentials %\nog{exhibit} 
are very short-ranged, 
the particles located at different trap wells can still 
interact due to the small overlaps between the corresponding
\emph{SW wave functions}
[cf.\ Fig.~\ref{fig:pot}(a)], i.\,e., the non-perfect localization of the Wannier functions in most realistic lattice systems. 
This finite overlap facilitates the control of atom-atom interactions via
FRs of atoms at different lattice sites, 
as previously observed, e.\,g., in the Mott-to-superfluid  transition~\cite{cold:grei02a, cold:bakr10}.

Figure~\ref{fig:pot} also sketches 
some of the models for the trapping potentials
that we have considered when modeling the experiment.
Figures~\ref{fig:pot}(b) and~\ref{fig:pot}(c)
show the 
SW
%single-well
description 
obtained by Taylor expanding
a $\sin^2$-like optical potential
%the external trapping potential
around the origin
%up to
%%second order
%order~2
%(yielding the harmonic trap, SW2, provided by Eq.~\eqref{eq:Vharm})
%and~6
%(sextic potential, SW6, provided by Eq.~\eqref{eq:Vsw})
%% (sextic potential, SW6)
as
\begin{eqnarray}
   \mathcal{V}_{\rm opt}(q_i) &=& V_q \sin^2 \left( k q_i \right )   \label{eq:Vsin2} \\
                &\approx &
%   V(q) = V_{q} \sin^2 \left[ ( 2\pi/\lambda) q \right ] \approx 
%   V(q) = V_{q} \sin^2 \left( \frac{ 2\pi}{\lambda} q \right ) \approx 
%V_q \left(
%k^2 q_i^2 - \frac{1}{3} k^4 q_i^4 +  \frac{1}{45} k^6 q_i^6
%\right)
% + \mathcal{O} \left( k^8 q_i^8 \right)  \!  \! , \, \\
%
V_q  ( k q_i)^2 + \mathcal{O} \left( (k q_i)^4 \right) ,  \label{eq:Vharm} \\
                &\approx &
V_q \left(
( k q_i)^2 - \frac{ (k q_i)^4 }{3} +  \frac{ 2 ( k q_i)^6}{45}
\right) %\nonumber \\
%&&
+ \, \mathcal{O} \left( (k q_i)^8 \right) ,  %\nonumber \\
%
%V_q \left(
%( k q_i)^2 - \frac{1}{3} (k q_i)^4 +  \frac{1}{45} ( k q_i)^6
%\right)
% + \mathcal{O} \left( (k q_i)^8 \right) , \nonumber \\
 \label{eq:Vsw}
\end{eqnarray}
with~$q_i=x_i, y_i$, and~$z_i$ the position coordinates of 
particle~$i=1,2$, $k=2\pi/\lambda$ the wave number, 
and~$V_q$ the potential depth (see discussion below).
As can be seen, the expansion of Eq.~\eqref{eq:Vsin2} 
up to order~2 yields the harmonic approximation 
provided by Eq.~\eqref{eq:Vharm},
while its expansion up to order~6 renders the sextic potential 
provided by Eq.~\eqref{eq:Vsw}.

Furthermore, in order to provide a more realistic description of 
the optical lattice, which also facilitates the study of off-site phenomena~\cite{cold:garc12, cold:garc14},
we also consider in the~$z$ direction a DW potential.
This model, shown in Figs.~\ref{fig:pot}(d) and~\ref{fig:pot}(e),
is obtained by Taylor expanding a $\cos^2$-like optical potential
around the origin up to order~12 as
\begin{eqnarray} 
   \mathcal{V}_{\rm opt}(z_i) &=& V_z \cos^2 \left( k z_i \right )  \label{eq:Vcos2}  \\
                &\approx &
V_z \left( 1- 
(k z_i)^2 + \frac{(k z_i)^4}{3}  - \frac{ 2 (k z_i)^6 }{45} + \frac{ (k z_i)^8 }{315} \right. 
-
\left.
 \frac{ 2 (k z_i)^{10} }{14175}  + \frac{2 (k z_i)^{12} }{467775}
\right) \nonumber \\
 && + \mathcal{O} \left( (k z_i)^{14} \right)  . %\nonumber  \\
\label{eq:Vdw}
%
%V_q \left( 1- 
%k^2 q_i^2 + \frac{1}{3} k^4 q_i^4 - \frac{1}{45} k^6 q_i^6 + \frac{1}{315} k^8 q_i^8 \right.  \nonumber \\
%&-& \left.
% \frac{1}{14175} k^{10} q_i^{10} - \frac{1}{467775} k^{12} q_i^{12}
%\right)
% + \mathcal{O} \left( k^{12} q_i^{14} \right)  .\nonumber  \\
\end{eqnarray}

To conclude this section, let us summarize
the models that have been considered here:
\begin{enumerate}[label=(\roman*)]
\item Single-well model of order 2 (SWM2). It constitutes a perfectly 
      harmonic (SW) approximation to the trapping potential in all 
      spatial directions [cf.\ Eq.~\eqref{eq:Vharm}].

\item Single-well model of order 6 (SWM6). It is an anharmonic 
      sextic description of the lattice in all directions,
      which also provides a more realistic SW description of 
      a lattice [cf.\ Eq.~\eqref{eq:Vsw}].

\item Double-well model (DWM). It combines a DW potential in the~$z$ 
      direction [cf.\ Eq.~\eqref{eq:Vdw}]  
      and two sextic potentials in the transversal directions 
      [cf.\ Eq.~\eqref{eq:Vsw}].

\end{enumerate}

Notice that, for simplicity, we refer to model (iii) as DWM,
despite its transversal confinement being described by 
two sextic (and not harmonic) potentials.
Additional calculations were also performed using a model 
combining double wells in two of the directions, and
the sextic potential in the remaining one, 
as well as a triple-well model (by Taylor expanding Eq.~\eqref{eq:Vsin2}
up to order~22, see \cite{cold:schn09}, in one direction
combined with two sextic expansions
in the remaining directions).
For the remainder of this work,
we omit these additional multi-well models
since their predictions are already captured by 
the model (iii) (DWM), which is carefully discussed, 
along with its relation to models (i) and (ii).

Besides the different models for the optical lattice,
different geometries (Q1D and 3D) and interaction descriptions
(\emph{ab-initio} curves and $\delta$ regularized pseudopotential~\cite{cold:blum02})
are considered. All the parameters in our simulations
(potential depths, wave length, atomic properties, etc.) 
correspond to the values
%\nog{known from} 
as reported in the experiment \cite{cold:lee23},
i.\,e., none of them have been fitted.

%==================================================================
\subsection{Numerical 
%\nog{scheme}
approach}
\label{sec:numerics}
%==================================================================

The study of Eq.~\eqref{eq:H} is very challenging,
even in the stationary case,
where one is essentially interested in
the computation of the system's eigenenergies and eigenstates.
The reason is twofold.
First, if the full dimensionality of the
lattice (3D) is considered,
six-dimensional interparticle-interaction integrals must be solved.
Second, the calculation of the matrix elements is also very demanding 
due to the dramatically different scales
of the short-range interatomic potential,
of the order of a few atomic units,
and the trapping potential,
which typically exceeds hundreds or even thousands of atomic units.

These challenges are mostly circumvented by
changing from absolute to CM and RM coordinates.
Then, the interaction potential solely depends on the RM 
coordinate~$\mathbf{r} = \mathbf{r}_1 - \mathbf{r}_2$. 
However, in RM and CM coordinates, 
the optical-trap potential depends, in general, on both 
coordinates, the CM one defined as
$\mathbf{R} = \mu_1  \mathbf{r}_1 + \mu_2  \mathbf{r}_2$,
with~$\mu_i = m_i /  (m_1+m_2)$ being
the reduced mass of particle~$i$. 
Nevertheless, if the trapping potential is expanded as,
e.\,g., in Eqs.~\eqref{eq:Vharm},~\eqref{eq:Vsw}, 
and~\eqref{eq:Vdw}, or it is fitted using other polynomial sets,
the total potential can be split as
\begin{eqnarray}
\label{eq:VCMrm}
   \hat{\mathcal{V}}_{\rm opt} \left( \mathbf{r}_1 \right) +   \hat{\mathcal{V}}_{\rm opt} \left( \mathbf{r}_2 \right)
    + \hat{U} \left( \mathbf{r}_1,  \mathbf{r}_2 \right)
%   \hat{\mathcal{V}}_{\rm opt}^{(1)} \left( \mathbf{r}_1 \right) +   \hat{\mathcal{V}}_{\rm opt}^{(2)} \left( \mathbf{r}_2 \right)
=  % \nonumber \\
\hat{\mathcal{V}}_{\rm CM} \left( \mathbf{R} \right) +
 \hat{\mathcal{V}}_{\rm RM} \left( \mathbf{r} \right) +
 \hat{\mathcal{W}}_{\rm CM-RM} \left( \mathbf{R},  \mathbf{r} \right),
\end{eqnarray}
where~$\hat{\mathcal{V}}_{\rm CM} \left( \mathbf{R} \right)$
and~$\hat{\mathcal{V}}_{\rm RM} \left( \mathbf{r} \right)$ are the CM and RM potential-energy operators,
and~$ \hat{\mathcal{W}}_{\rm CM-RM} \left( \mathbf{R},  \mathbf{r} \right)$ is a term that accounts
for the CM-RM coupling due to, e.\,g., anharmonicities of the 
potential or a possible mass imbalance. Then, Eq.~\eqref{eq:H} can 
be readily recast as 
\begin{equation}
\label{eq:HCMrm}
   \hat{\mathcal{H}} \left( \mathbf{r},  \mathbf{R} \right)
=   \hat{{h}}_{\rm CM} \left( \mathbf{R} \right) +
 \hat{{h}}_{\rm RM} \left( \mathbf{r} \right) +
 \hat{\mathcal{W}}_{\rm CM-RM} \left( \mathbf{R},  \mathbf{r} \right) ,
\end{equation}
where
\begin{equation}
\label{eq:HCM}
 \hat{{h}}_{\rm CM} \left( \mathbf{R} \right) =
-\frac{\hbar^2}{2M} \hat{\mathbf{P}}_{\rm CM}^2 + \hat{\mathcal{V}}_{\rm CM}\left( \mathbf{R} \right) 
\end{equation}
is the CM Hamiltonian, 
and
\begin{equation}
\label{eq:Hrm}
 \hat{{h}}_{\rm rm} \left( \mathbf{r} \right) =
-\frac{\hbar^2}{2\mu} \hat{\mathbf{p}}_{\rm RM}^2 + \hat{\mathcal{V}}_{\rm RM}\left( \mathbf{r} \right) 
\end{equation}
is the RM Hamiltonian, with the reduced mass $\mu=m_1 m_2 / (m_1+m_2)$.

For systems of equal masses under perfectly harmonic confinement,
there is no coupling between the CM and the RM
(${\mathcal{W}}_{\rm CM-RM} = 0$).
In this idealized scenario,
an analytical expression for the eigenenergies of Eq.~\eqref{eq:H}
exists if the interatomic interaction is modeled using a 
contact $\delta$ pseudopotential~\cite{cold:busc98, cold:blum02}.

However, numerical approaches are essential to accurately compute the system's eigenenergies for other types of trapping potentials --- such 
as those considered in this work. Similarly, if more general interatomic interactions are introduced, including Lennard-Jones, Morse, or realistic
\emph{ab-initio} potentials (as employed here),
the problem becomes significantly more challenging,
%\nog{complex},
so that it
%\nog{the problem}
can be only numerically solved since 
analytical solutions are no longer feasible.

In this work, we calculate the eigenenergies and the corresponding eigenfunctions for the Hamiltonian in Eq.~\eqref{eq:HCMrm} using
the method introduced in the Ref.~\cite{cold:gris09},
which has been successfully applied to study ultracold atomic systems
trapped in a large variety of
sinusoidal-like optical lattices~\cite{cold:gris11, cold:sala12,cold:sala13, cold:cape23}. 
If the trapping potential is symmetric with respect to the origin
and separable in the different directions,
then the Hamiltonian~\eqref{eq:H}
belongs to the~$\mathcal{D}_{2h}$ 
point-symmetry group~\cite{gen:rotman95}.
As a consequence, the system eigenfunctions satisfy the symmetries
of the corresponding eight irreducible representations (IRs). The 
method first constructs the Hamiltonian matrices
associated with Eqs.~\eqref{eq:HCM} and~\eqref{eq:Hrm}
in a  basis set formed by $\mathcal{B}$ splines and 
spherical harmonics~$Y_m^\ell$ 
(or combinations of them in order to account for the symmetry
of a particular IR of interest).
The diagonalization of these matrices yields the 
eigenenergies and eigenfunctions for the CM~$\psi_i^{\rm CM}(\mathbf{R})$
and RM~$\varphi_j^{\rm RM}(\mathbf{r})$.
Subsequently, a novel basis set formed of \emph{configurations}
constructed as products
$\Psi_{i,j} (\mathbf{R}, \mathbf{r}) = \psi_i^{\rm CM}(\mathbf{R}) \varphi_j^{\rm RM}(\mathbf{r})$
of the previous single-coordinate wavefunctions is created. 
Finally, the Hamiltonian matrix~\eqref{eq:HCMrm} is computed
in the~$\Psi_{i,j}$ full configuration-interaction %(CI)
basis set, and eventually diagonalized,
this rendering the eigenenergies and eigenfunctions
for the complete two-particle system.\footnote{
%---------------------------------------
%BASIS SET DESCRIPTION
%---------------------------------------
In particular,
the computation of the eigenenergies and eigenfunctions
associated with the CM Hamiltonian~\eqref{eq:HCM}
%(see Eq.~(\ref{eq:HCM}))
has been performed using 
102  $\mathcal{B}$ splines linearly distributed 
up to a distance of $\sqrt 3 \times 10^4 a_0$,
along with linear superpositions
of spherical harmonics, 
$\mathcal{Y}_{\ell, m}^\pm  = Y^\ell_m \pm Y^{-\ell}_m$,
with~$\ell =0, 1, 2, \ldots, 34$, and~$m =0, 2, \ldots, 8$.
%with~$l =0, 2, 4, \ldots, 34$, and~$m =0, 2, \ldots, 8$.
When only $\mathcal{Y}_{\ell, m}^+$
combinations and even~$\ell$ and~$m$ values
%in $\mathcal{Y}_{\ell, m}^+$
are considered, the totally symmetric states that belong to 
the $a_g$ IR~\cite{gen:rotman95} are computed,
while the consideration of only odd~$\ell$
and even~$m$ values yield
states that are antisymmetric with respect to the~$xy$ plane
and symmetric with respect to the remaining ones,
which belong to the $b_{1u}$ IR. %
The $a_g$ eigenenergies and eigenfunctions
related to the RM Hamiltonian~\eqref{eq:Hrm}
%(see Eq.~(\ref{eq:Hrm})),
are obtained with
the same combinations of spherical harmonics,
%have been considered,
but 152 $\mathcal{B}$ splines
in order to adequately describe the short-range
interatomic potential,
half of them linearly distributed
up to a distance of~15$a_0$, with the Bohr radius $a_0$, 
and the other half following a geometric distribution
so their separation increases with the distance
up to a distance of $\sqrt{12} \times 10^4 a_0$.
Then,
the full Hamiltonian is subsequently diagonalized
in a basis set that combines 
for the Q1D optical trap ($V_z = 10.7 E_R$)
80 CM eigenstates with 200 RM eigenstates,
and
for the 3D trap ($V_z = 35 E_R$)
120 CM eigenstates with 300 RM eigenstates.
%----------------------------------------------
% END BASIS SET DESCRIPTION
%----------------------------------------------
}

%==================================================================
% RESULTS
\section{Results and discussion} \label{sec:results}
%==================================================================
%\subsection{Results} \label{sec:results}
%==================================================================
In this section,
we present the main findings of our study
along with the corresponding discussion.
This section is divided into two parts. 
Section~\ref{sec:results_multi} contains the results for
the DWM introduced in Sec.~\ref{sec:models}. 
Section~\ref{sec:results_sw} discusses the performance of SW models
(both harmonic, SWM2, and anharmonic, SWM6),
paying special attention to an apparent
symmetry violation, which does not appear
%\nog{is not present} 
in the DWM. 

%\newpage
%==================================================================
\subsection{Multiwell effects on
                        inelastic confinement-induced resonances} \label{sec:results_multi}
%==================================================================

%==================================================================
\subsubsection{Quasi-one dimensional confinement} \label{sec:results_multi_q1d}
%==================================================================

\begin{figure}[hbtp!]
\includegraphics[width=0.95\columnwidth]{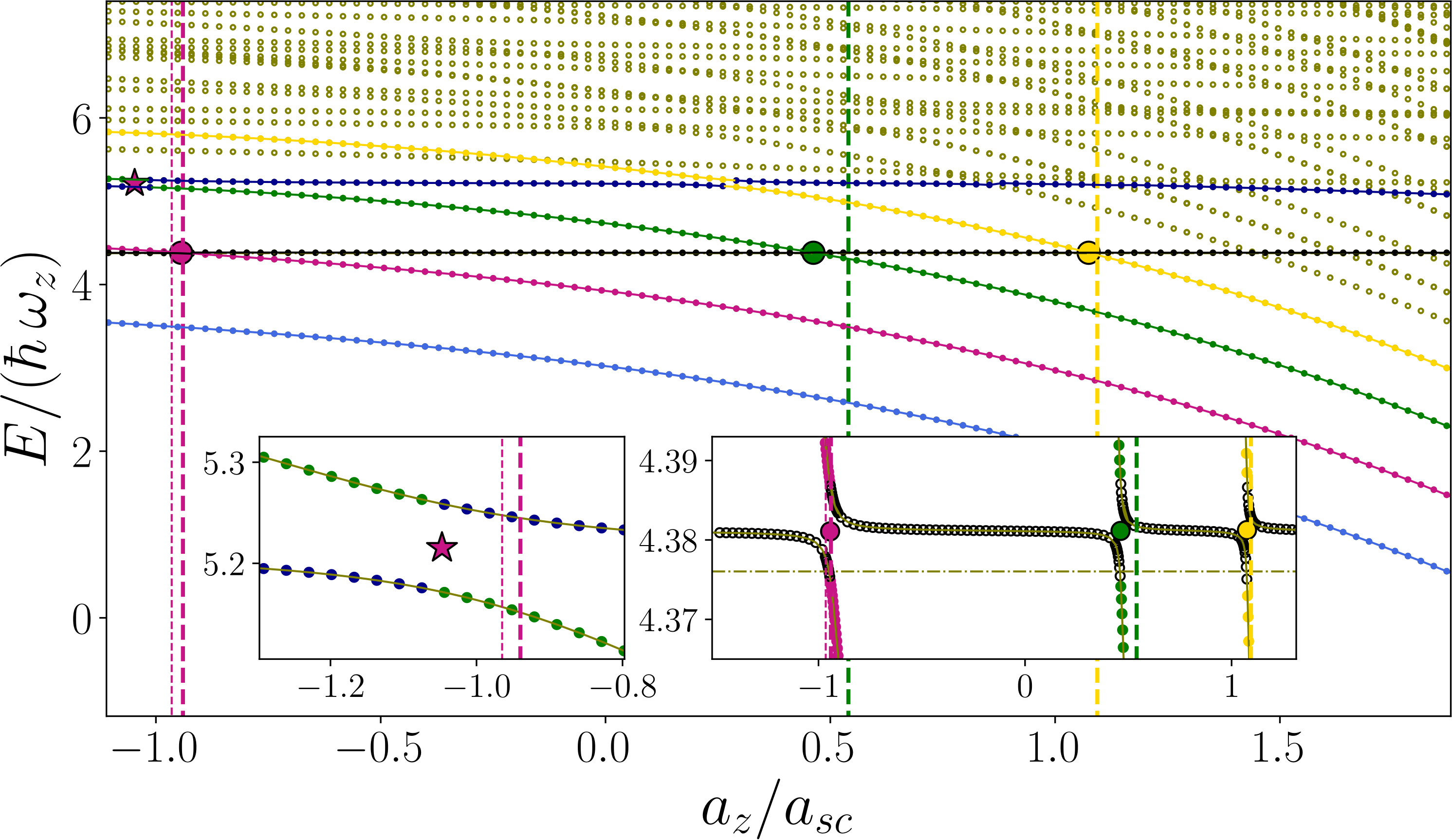}
%aniso_dw_4300-4300-1315_q1d_colored_mod_tex-fonts_top_icir.pdf}
%
%\includegraphics[width=0.95\columnwidth]{figs/aniso_dw_4300-4300-4300-1315_q1d-3d_colored_mod}
%\includegraphics[width=0.95\columnwidth]{figs/aniso_dw_4300-4300-4300-1315_q1d-3d_colored2}
%\includegraphics[width=0.95\columnwidth]{figs/aniso_dw_4300-4300-4300-1315_q1d-3d_colored1}
%
%\includegraphics[width=0.95\columnwidth]{figs/spectrum1a_iso_dw_4300-4300-4300.png}
%\includegraphics[width=0.95\columnwidth]{figs/spectrum1b_aniso_dw_4300-4300-1315}
\caption{Energy spectra for 
totally symmetric eigenstates ($A_g$ irreducible representation) of
two ultracold
%\nog{Li$^7_2$}
$^7$Li atoms  
confined in an optical lattice modeled with the sextic trap~\eqref{eq:Vsw}
in the~$x$ and~$y$ directions,
and a double well~\eqref{eq:Vdw} in the~$z$ direction
[model (iii)--DWM]
as a function of the ratio between the characteristic
trap size~$a_z= \sqrt{2 \hbar/(m_{\rm Li} \omega_z)}$, 
and the scattering length~$a_{\rm sc}$.
The trap depth equals~$35\,E_R$ in~$x$ and~$y$ directions,
and~$10.7\,E_R$, in~$z$ direction.
Highlighted branches correspond to the states listed in the
Table~\ref{tab:states_vs_colors}.
Vertical dashed lines indicate the positions of the 
dispersive features reported in the experiment in Ref.~\cite{cold:lee23}. Insets: enlarged view of some of the avoided crossings.
}
\label{fig:spectra_dw_q1d}
%\label{fig:spectra_dw}
\end{figure}

For the case of a Q1D trap,
Fig.~\ref{fig:spectra_dw_q1d} shows the energy spectra
of the totally symmetric eigenstates of Eq.~\eqref{eq:H}, 
i.\,e., the ones
%\nog{which}
belonging to the $A_g$ IR~\cite{gen:rotman95}.
%\nog{Figure~\ref{fig:spectra_dw_q1d} shows the results
%Figure~\ref{fig:spectra_dw}(a) shows the results
%for the Q1D trap.} 
%and Fig.~\ref{fig:spectra_dw}(b) for the 3D trap. In both cases, 
The trapping potential is described by 
the DWM, i.e.,
%the DW model, i.e.,
by a DW in the $z$ direction and
the SW as defined in Eq.~\eqref{eq:Vsw}
(sextic trap)
in the~$x$ and~$y$ directions.
%The insets present enlarged views of the lowest-lying avoided crossings
%(see discussion below).
%Furthermore,
%The states of interest have been highlighted with different colors.
%
%--------------------------------------------------------------------------------------------
% Figure: Q1D SPECTRA
%------------------------------------------------------------.--------------------------------
%--------------------------------------------------------------------------------------
%
As can be seen, the energy spectrum 
comprises two series of branches 
with two different slopes as a function of the $a_z/a_{\rm sc}$ ratio,
depending on their underlying RM contributions,
with $a_z = \sqrt{2 \hbar /(m_{\rm Li} \omega_z)}$ the characteristic trap size.
The states of interest for this work are highlighted with 
the same colors
as those considered in the
Ref.~\cite{cold:lee23},
%\nob{different colors,}
%\nog{as indicated in} 
see
Table~\ref{tab:states_vs_colors}.
%
%---------------------------------------
% Table
%---------------------------------------
%
\begin{center}
\begin{table}[h]
  \caption{
    Colors associated with the eigenstates of the double-well model DWM (iii) and the corresponding ones for the single-well models, both harmonic SWM2 (i) and sextic SWM6 (ii),
    together with the corresponding relative‑motion (RM) kind 
    of state (molecular or trap). 
    The state marked with “$\times$” does not exist,
    while those marked with “$-$” are not shown or highlighted.
   These conventions are used throughout Figs.~\ref{fig:spectra_dw_q1d}–\ref{fig:icir_positions} 
    (in Figs.~\ref{fig:spectra_harm-sw_q1d} 
    and~\ref{fig:spectra_harm-sw_3d}, the results for the
    DWM are superimposed as solid olive lines).
}
  \label{tab:states_vs_colors}
  \footnotesize
  \begin{tabular}{lcccccccc}   % 2 columns, both centered
    \hline
    \multirow{2}{*}{RM kind} & \multicolumn{2}{c}{DWM} && \multirow{2}{*}{Fig.} & \multicolumn{2}{c}{SWM2 \& SWM6} & \multirow{2}{*}{Fig.} &
    \multirow{2}{*}{Color} \\
    \cline{2-3} \cline{6-7}
            & RM state & CM state    &&& RM state & CM state    &       \\
    \hline    \hline
    Molecular & 
    $\varphi^{\rm RM, LBS}_{{\rm DW}, a_g} $ &
    $\psi^{\rm CM}_{{\rm DW}, a_g}(0, 0, 0)$ && \ref{fig:wf_rmlbs-CM0-3}(a) &
    $\varphi^{\rm RM, LBS}_{{\rm SW}, a_g} $ &
     $\psi^{\rm CM}_{{\rm SW}, a_g}(0, 0, 0)$ & \ref{fig:wf_rmlbs-CM0-3}(a') &Blue   \\
    & 
    $\varphi^{\rm RM, LBS}_{{\rm DW}, a_g} $ &
    $\psi^{\rm CM}_{{\rm DW}, a_g}(0, 0, 2)$ && \ref{fig:wf_rmlbs-CM0-3}(b) &
    $\varphi^{\rm RM, LBS}_{{\rm SW}, a_g} $ &
     $\psi^{\rm CM}_{{\rm SW}, b_{1u}}(0, 0, 1)$ & \ref{fig:wf_rmlbs-CM0-3}(b') &Magenta   \\
    & 
    $\varphi^{\rm RM, LBS}_{{\rm DW}, a_g} $ &
    $\psi^{\rm CM}_{{\rm DW}, a_g}(0, 0, 4)$ && \ref{fig:wf_rmlbs-CM0-3}(c) &
    $\varphi^{\rm RM, LBS}_{{\rm SW}, a_g} $ &
     $\psi^{\rm CM}_{{\rm SW}, a_g}(0, 0, 2)$ & \ref{fig:wf_rmlbs-CM0-3}(c') &Green   \\
    & 
    $\varphi^{\rm RM, LBS}_{{\rm DW}, a_g} $ &
    $\psi^{\rm CM}_{{\rm DW}, a_g}(0, 0, 6)$ && \ref{fig:wf_rmlbs-CM0-3}(d) &
    $\varphi^{\rm RM, LBS}_{{\rm SW}, a_g} $ &
     $\psi^{\rm CM}_{{\rm SW}, b_{1u}}(0, 0, 3) $ & \ref{fig:wf_rmlbs-CM0-3}(d') &Yellow   \\
    & 
    $-$ & $-$ && $-$ &
    $\varphi^{\rm RM, LBS}_{{\rm SW}, a_g} $ &
     $\psi^{\rm CM}_{{\rm SW}, b_{1u}}(2, 0, 1) $ &  $-$ &Brown   \\
    & 
    $-$ & $-$ && $-$ &
    $\varphi^{\rm RM, LBS}_{{\rm SW}, a_g} $ &
     $\psi^{\rm CM}_{{\rm SW}, b_{1u}}(0, 2, 1) $ &  $-$ &Brown   \\
    \hline

    Trap & 
    $\varphi^{\rm RM, TS1}_{{\rm DW}, a_g} $ &
    $\psi^{\rm CM}_{{\rm DW}, a_g}(0, 0, 0)$ && \ref{fig:wf_mott}(a) &
    $\times$ &
    $\times$ & $\times$ & Black   \\
    & 
    $\varphi^{\rm RM, TS2}_{{\rm DW}, a_g} $ &
    $\psi^{\rm CM}_{{\rm DW}, a_g}(0, 0, 0)$ && \ref{fig:wf_mott}(b) &
    $\varphi^{\rm RM, TS1}_{{\rm SW}, a_g} $ &
    $\psi^{\rm CM}_{{\rm SW}, a_g}(0, 0, 0)$ & \ref{fig:wf_mott}(a') & Dark blue    \\
    & 
     $-$ & $-$ && $-$ &
%    $\varphi^{\rm RM, TS1}_{{\rm DW}, a_g} $ &
%    $\psi^{\rm CM}_{{\rm DW}, a_g}(0, 0, 0)$ &&
    $\varphi^{\rm RM, TS1}_{{\rm SW}, a_g} $ &
     $\psi^{\rm CM}_{{\rm SW}, b_{1u}}(0, 0, 1)$ &  \ref{fig:wf_mott}(b') &Dark blue   \\
    \hline
  \end{tabular}
\end{table}
\end{center}
%\end{widetext}
%---------------------------------------
%

\begin{figure}[hbtp!]
\includegraphics[width=0.75\columnwidth]{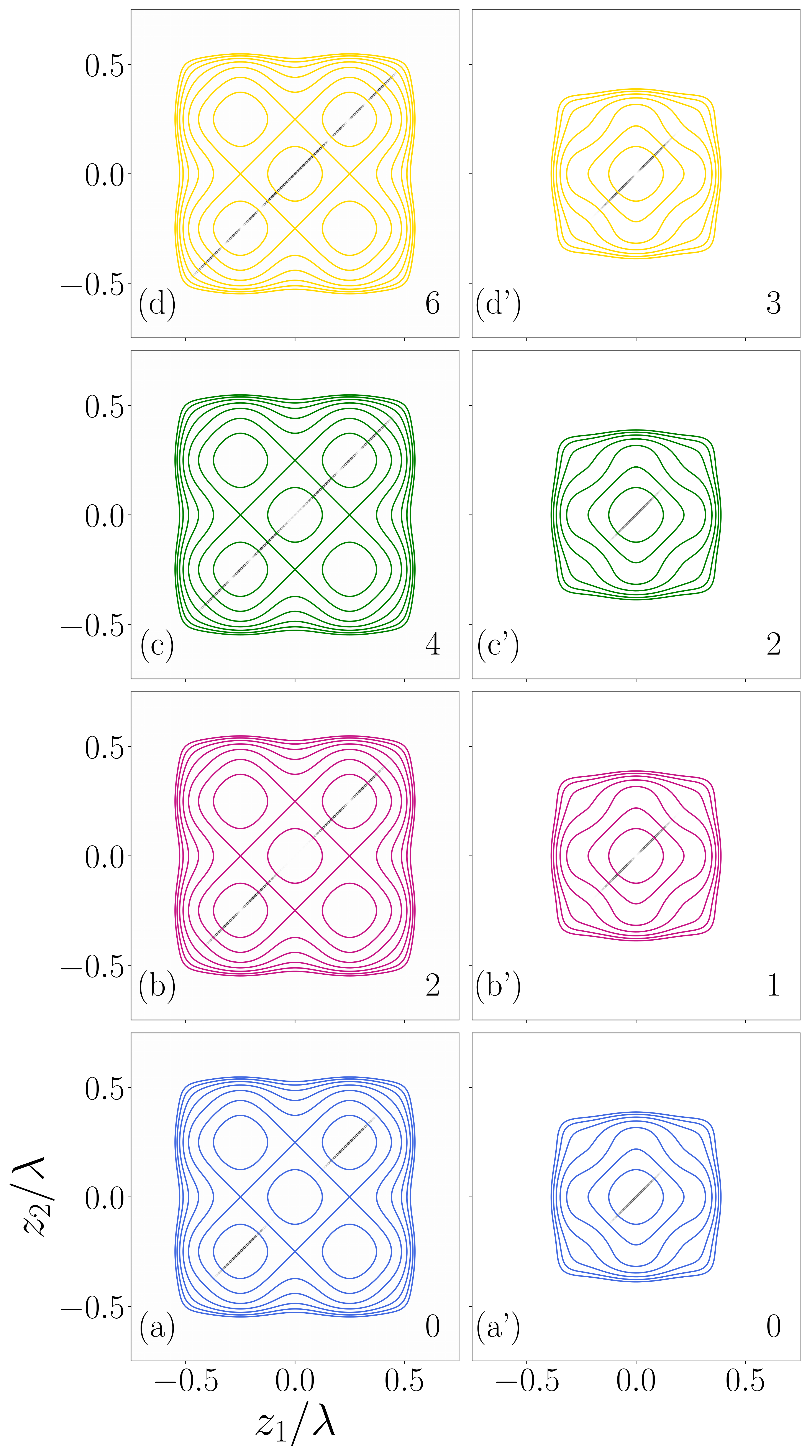}
%qqq
%\includegraphics[width=0.95\columnwidth]{figs/wf3d_RMlbs_swCM0-3_dwCM0-6_opc1}
\caption{
Cuts along the $z_1,z_2$ plane of the
probability densities through the
least-bound molecular states
(gray elongated lobes)
%along~$z_2 = z_1$)}
%with~$n_{\rm DW}^{\rm CM} = 0, 1, 2,$ and~3
for a quasi‑one‑dimensional optical trap 
($V_x = V_y = 35\,E_R$, $V_z = 10.7\,E_R$) 
(a) to (d) for the double-well potential and thus the states 
$\varphi^{\rm RM, LBS}_{{\rm DW}, a_g}  
\psi^{\rm CM}_{{\rm DW}, a_g}(0, 0, n_{\rm DW}^{\rm CM})$
and (a') to (d')
for the sextic single-well potential and the states
$\varphi^{\rm RM, LBS}_{{\rm SW6}, a_g}  \psi^{\rm CM}_{{\rm SW6}, a_g/b_{1u}}(0, 0, n_{\rm SW6}^{\rm CM})$,
with
$n_{\rm DW}^{\rm CM}$ and $n_{\rm SW6}^{\rm CM}$
specified in the lower right corner of each panel.
%The number in the right corner of each panel gives the
%corresponding quantum number (left) $n_{\rm DW}^{\rm CM}$, and (right) $n_{\rm SW6}^{\rm CM}$.
The lines mark the equipotential lines for~$V(z_1, z_2)/V_z$
proportional to 0.5 using the same colors 
as listed in the Table~\ref{tab:states_vs_colors}.
%
% Q1D
% wz (Ix=1315kW/cm^2)  :  2.4979600481639766e-11 {\au} = 164.357873904878 kHz
% (w is)
% ah (Ix=1315nkW/cm^2) :  2502.057536391109      {\au}
% 
% 3D
% wz (Ix=4300kW/cm^2) : 4.517070194312925e-11 {\au} = 297.208938134159 kHz
% ah (Ix=4300kW/cm^2) : 1860.636844561505    {\au}
%
% asc (\delta = 0.70650) =   990.720275990685 {\au} 
% asc (\delta = 0.73000) =  -529.765832889192  {\au}
%
% ah(Q1D) / asc (\delta = 0.70650) =  2.52549341830028
% ah(Q1D) / asc (\delta = 0.73000) = -4.72294999234964
%
% ah (Ix=1315nkW/cm^2) / asc (\delta = 0.70650) = 2.5255
%
% TwoAt InOL: w_1kHz = 1.519829846003E-13  ! in {\au}
%                                              This factor includes a 2*\pi factor (*1000 Hz/kHz), since
%
%                                               1{\au} = 4.1341373335182E13 kHz,
%
%                                                and then
%
%                                                1kHz = 1/ 4.1341373335182E13 
%                                                           = 2.41888432658571e-14 
%                                                           = ( 2 * pi ) * w_1kHz .
%
%\fr{New labels (a), (a'), (b), (b')...}
%\CG{[AS: comment on color coding]}
\label{fig:wf_rmlbs-CM0-3}
}
\end{figure}

On the one hand, the branches with the most pronounced dependence of the energy 
on the $a_z/a_{\rm sc}$ ratio 
correspond to {\it molecular} states
%the largest slopes
%correspond to molecular dimers
%\nog{that interact in a strong fashion} 
showing 
%\nog{thus}
a strong influence of the
interatomic interaction and thus on $a_{\rm sc}$.
The corresponding total 
%\nog{eigenfunctions}
wavefunctions
%\fr{``eigenfunctions'' is more specific than ``wavefunctions''}
combine the RM least bound state~$\varphi_{\textnormal{DW}, a_g}^{\rm RM, LBS}$,
with different CM states
%$\psi_j^{\rm CM} =
$\psi_{\textnormal{DW}, a_g}^{\rm CM}(n_x^{\rm CM}, n_y^{\rm CM}, n_z^{\rm CM})$.
%\nog{Some examples for}
Cuts through the probability densities of the lowest-lying states 
%of the Q1D model
are shown in Fig.~\ref{fig:wf_rmlbs-CM0-3}.
As can be seen,
in all cases the probability density
%\nog{localizes} 
in this plane is mostly localized
along the diagonal $z_2=z_1$, which also corresponds to the CM axis
(where the RM coordinate is~$z=0$),
%which corresponds to a small RM coordinate,
as expected for a molecular state.
Notice, also, that the number of nodes along that diagonal
%,
%which also corresponds to the CM axis,
increases
from the bottom to the top
if the CM excitation in 
the $z$ direction,
$n_z^{\rm CM}$, changes from
0 [Fig.~\ref{fig:wf_rmlbs-CM0-3}(a)],
to~2 [Fig.~\ref{fig:wf_rmlbs-CM0-3}(b)],
4 [Fig.~\ref{fig:wf_rmlbs-CM0-3}(c)],
%\nob{to}
and
6 [Fig.~\ref{fig:wf_rmlbs-CM0-3}(d)];
while, in these cases,
%all presented cases,
$n_x^{\rm CM} = n_y^{\rm CM} = 0$ holds.
%
%--------------------------------------------------------------------------------------------
% Figure: WF FOR THE STATES THAT RENDER ICIRs IN QUASI-1D
%  RM: LEAST-BOUND MOLECULAR STATE + CM0-3
%------------------------------------------------------------.--------------------------------
%--------------------------------------------------------------------------------------
%

%
%--------------------------------------------------------------------------------------------
% Figure: WF FOR THE DW GROUND STATE (MOTT INSULATOR)
%------------------------------------------------------------.--------------------------------
\begin{figure}[hbtp!]
\includegraphics[width=0.75\columnwidth]{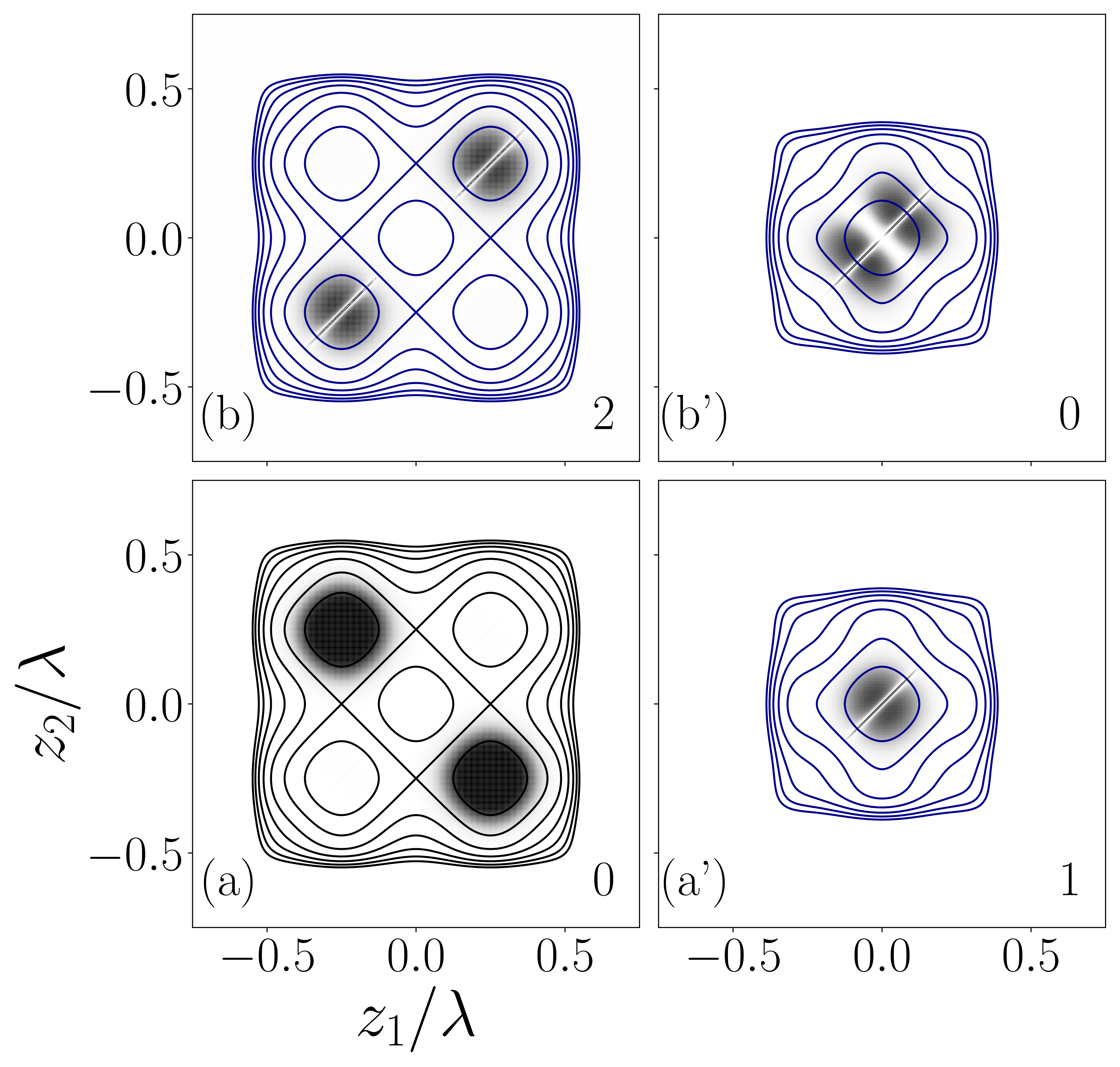}
\caption{Same as Fig.~\ref{fig:wf_rmlbs-CM0-3}, but for
(a) the first and
(b) the second trap states of the double-well model, 
%\nog{and}
as well as for (a') the first
$A_g$ and (b')~$B_{1u}$ trap states of the sextic single-well 
model (cf. Sec.~\ref{sec:models}).
}
%
% Q1D
% wz (Ix=1315kW/cm^2)  :  2.4979600481639766e-11 {\au}
% ah (Ix=1315nkW/cm^2) :  2502.057536391109      {\au}
% 
% 3D
% wz (Ix=4300kW/cm^2) : 4.517070194312925e-11 {\au}
% ah (Ix=4300kW/cm^2) : 1860.636844561505    {\au}
%
% asc (\delta = 0.70650) =   990.720275990685 {\au} 
% asc (\delta = 0.73000) =  -529.765832889192  {\au}
%
% ah(Q1D) / asc (\delta = 0.70650) =  2.52549341830028
% ah(Q1D) / asc (\delta = 0.73000) = -4.72294999234964
%
\label{fig:wf_mott}
\end{figure}
%--------------------------------------------------------------------------------------
%

On the other hand, the almost horizontal curves
in Fig.~\ref{fig:spectra_dw_q1d},
whose energies are almost unaffected 
by the $a_z/a_{\rm sc}$ ratio, 
%the smallest slopes
are associated with
RM \emph{trap states}. 
In the absence of a trap they would correspond 
to two free (unbound) atoms.
%where the interaction between the two atoms %is less substantial,
%where the two atoms
%do not interact so strongly,
%are 
%in a RM state~$\varphi_{\textnormal{DW}, j}^{\rm RM, TS}$
%associated with a situation where the two atoms are
%separated in the RM coordinate,
%and then changes in the $a_z/a_{\rm sc}$ ratio influence much less the energy.
%\nog{Furthermore, like}
As in the case of the molecular states,
the CM part $\psi^{\rm CM}(n_x^{\rm CM}, n_y^{\rm CM}, n_z^{\rm CM})$
%\nog{may present} 
allows for different numbers of CM excitations.
In the figure, 
the first-trap state
$\varphi^{\rm RM, TS1}_{{\rm DW}, a_g}  \psi^{\rm CM}_{{\rm DW}, a_g}(0, 0, 0)$
and the second trap state
$\varphi^{\rm RM, TS2}_{{\rm DW}, a_g}  \psi^{\rm CM}_{{\rm DW}, a_g}(0, 0, 0)$
have been marked in black and dark blue colors,
respectively.
The probability densities of
these two states are presented in
Fig.~\ref{fig:wf_mott}.
%Fig.~\ref{fig:wf_mott}(a) and~\ref{fig:wf_mott}(c).
As can be seen,
the (lowest-lying) first trap state shown in Fig.~\ref{fig:wf_mott}(a)
has most of its probability density
along the off-diagonal~$z_2=-\,z_1$
and thus at different wells,
which is associated with a large RM coordinate, 
i.\,e., a configuration where both atoms are located at different wells.
The probability density along the diagonal~$z_2=z_1$ is, 
on the contrary, comparatively negligible.
Conversely, the maxima of 
the probability density of
the second trap state is primarily 
concentrated 
%\nog{along}
close to the diagonal~$z_2=z_1$,
which corresponds to a small value of the RM coordinate, 
i.\,e., a configuration where both atoms are located at the same well. 
However, in contrast to the bound molecular states, this state 
is  with respect to RM
an antibonding one,
as can be seen from the 
node along the diagonal $z_1=z_2$.
This different configuration also manifests in the slope of the corresponding
branches. Note, while
%as
the second trap state is moderately influenced by variations in the~$a_z/a_{\rm sc}$ ratio
(a behavior that is also observed in the SW model,
see discussion in Sec.~\ref{sec:results_sw}),
%while
the energy of the first trap state remains almost totally constant within
the considered range
(in fact, even in the non-interacting limits,
where~$a_z/a_{\rm sc} \rightarrow \pm \infty$, too).
As a consequence,
the first trap state
%This state 
of the DW potential plays the role of a perfect
Mott insulator with practically no interactions~\cite{cold:grei02a, cold:syas07, cold:robe08, cold:bakr10}
and single-site occupancy in an (ideal) infinite optical lattice
[see sketch in Fig.~\ref{fig:pot}(a)].

Due to symmetry conservation,
%\nog{when} 
if the interaction,
i.\,e., $a_{\rm sc}$, or the trap size,
i.\,e., $a_z$, are tuned and the
energy of a molecular state reaches that of a trap state
with the same symmetry
(in this case the $A_g$ IR),
an avoided crossing emerges in the spectrum~\cite{cold:vonNeum29}. 
This is associated with an ICIR,
because at such an avoided crossing an unbound (trapped) 
atom pair can be transferred into a molecular bound state 
and {\it vice versa}.
The right inset of Fig.~\ref{fig:spectra_dw_q1d}
shows enlarged views of the avoided crossings
that involve the first trap state,
\emph{aka} Mott insulator, and the lowest-lying molecular 
states of interest. All these avoided crossings
are quite broad with widths 
of order~$\mathcal{O} \left( 10^{-2} \,\hbar \, \omega_z \right)$. 
The molecular states
in Fig.~\ref{fig:spectra_dw_q1d} shown in different colors are 
$\varphi^{\rm RM, LBS}_{{\rm DW}, a_g}  \psi^{\rm CM}_{{\rm DW}, a_g}(0, 0, 2)$ (magenta),
$\varphi^{\rm RM, LBS}_{{\rm DW}, a_g}  \psi^{\rm CM}_{{\rm DW}, a_g}(0, 0, 4)$ (green), and
$\varphi^{\rm RM, LBS}_{{\rm DW}, a_g}  \psi^{\rm CM}_{{\rm DW}, a_g}(0, 0, 6)$ (yellow). As they are non-degenerate,
each one is responsible for the emergence of a separate ICIR.

The vertical colored lines in Fig.~\ref{fig:spectra_dw_q1d}
mark the positions of the
dispersive features observed in the experiment 
described by Lee \emph{et al.}~\cite{cold:lee23}.
Remarkably, the positions  of the avoided crossings 
(clearly visible in the insets), i.\,e., of the ICIRs in the DWM,
show 
a very good  agreement with the experimental results. 
Recall that the energy spectrum has been obtained using only
the given experimental parameters, 
such as the laser intensity and an \emph{ab-initio} atom–atom 
interaction, without introducing any fitting procedure or
additional approximations beyond those present in the experiment 
itself.

Finally, notice the existence of a very wide avoided crossing
(magenta star), with an $a_z/a_{\rm sc}$ ratio relatively
close to the dispersive feature related to the narrow FR
(leftmost thin dashed line) (see Sec.~\ref{sec:experiment}).
As can be well seen in the enlarged view of the left inset 
of the figure, this avoided crossing is substantially wider 
($\sim7$ times wider) than the other ones discussed.
This avoided crossing corresponds to an ICIR due to the interaction
between the molecular state
$\varphi^{\rm RM, LBS}_{{\rm DW}, a_g}  
 \psi^{\rm CM}_{{\rm DW}, a_g}(0, 0, 4)$
and the \emph{second} trap state
$\varphi^{\rm RM, TS2}_{{\rm DW}, a_g}  \psi^{\rm CM}_{{\rm DW}, a_g}(0, 0, 0)$. 
While the large width should, in principle, lead to a good experimental visibility of this state, such a detection would certainly require initial population of the second 
trap state in which, as discussed above, both atoms are predominantly located in 
the same well which is also responsible for the large width. Such a population of the second trap state may occur due to thermal excitations, an imperfect preparation of the Mott state \cite{cold:cape23}, or diabatic effects due to a fast ramping of the magnetic field with which the scattering length is tuned. 
The study of this avoided crossing, i.\,e., ICIR, could be performed 
in more detail in an experiment starting with a Mott insulator 
with double occupancy on each site~\cite{cold:robe08},
or, even better, with a tweezer array  
that would enable
a more exquisite control of
the atoms' positions~\cite{cold:murm15, cold:hod20, cold:rutt23}.

%==================================================================
\subsubsection{Three-dimensional confinement} \label{sec:results_multi_3d}
%==================================================================

\begin{figure}[hbtp!]
\includegraphics[width=0.95\columnwidth]{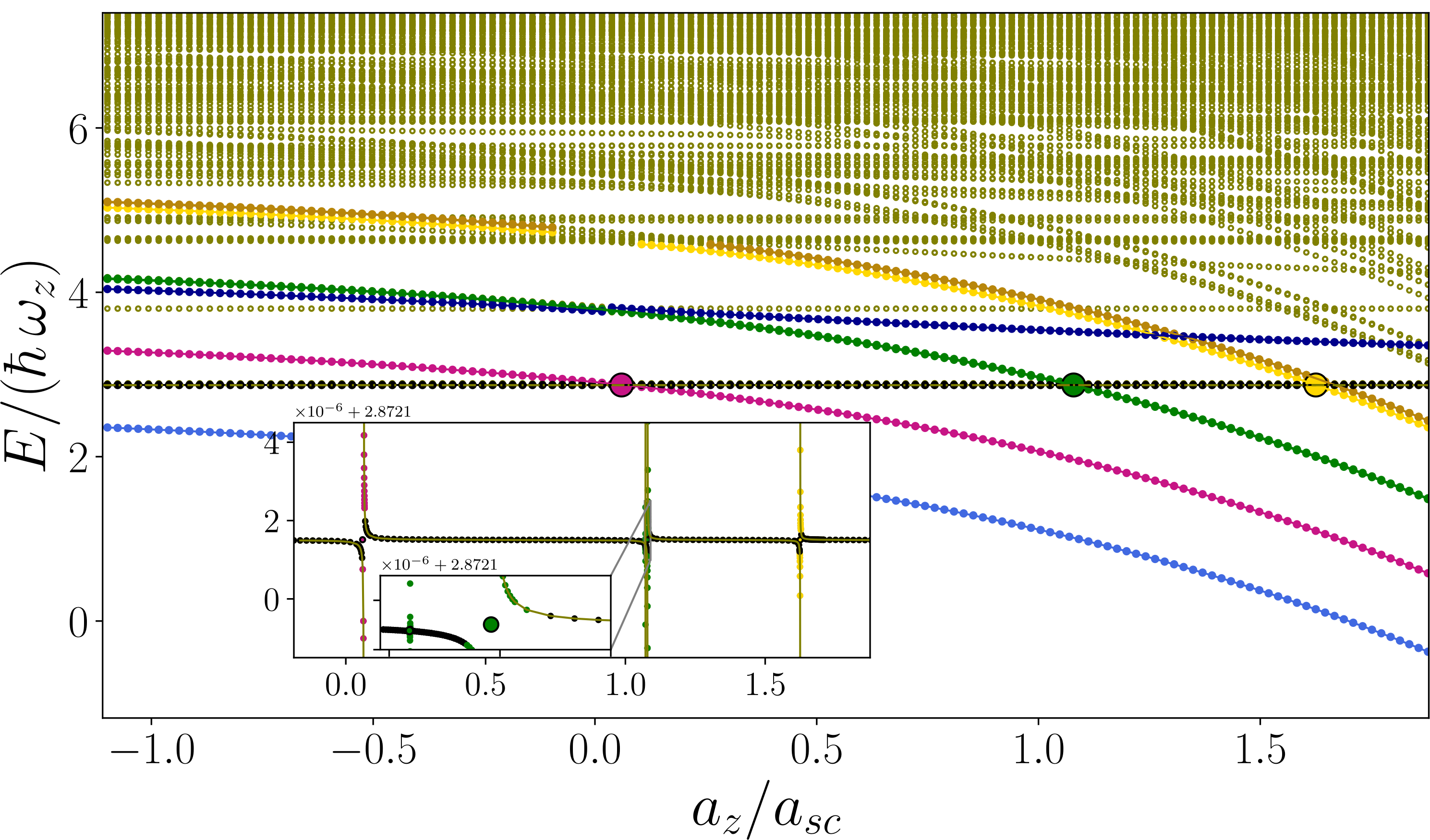}
%aniso_dw_4300-4300-4300_3d_colored_mod_tex-fonts_top_icir.pdf}
%
\caption{
Same as Fig.~\ref{fig:spectra_dw_q1d}, but for a 
trap with a depth equal to $35\,E_R$ in all directions.
}
\label{fig:spectra_dw_3d}
\end{figure}

Figure~\ref{fig:spectra_dw_3d} 
shows the results for a deep 3D 
%\nom{isotropic}
optical trap 
(1st experimental setting, 
$V_x=V_y=V_z$, see Sec.~\ref{sec:experiment}) described by
the DWM considered in Sec.~\ref{sec:results_multi_q1d},
%\nog{As i}
where the positions of three avoided crossings
%involving the
%Mott insulator state
have been hightlighted.
%These avoided crossings involve molecular states and
%the first trap state (Mott insulator).

As in the Q1D setup,
%\nob{In that case,}
the branches
which show a strong dependence on the $a_z/a_{\rm sc}$ ratio
%\nob{with large slopes}
correspond to molecular
states, while those with 
a 
%\nog{lesser}
smaller dependence
%\nob{smaller slopes}
are associated with 
trap states (see Table~\ref{tab:states_vs_colors}).
However, unlike in the Q1D setup, several states shown in Fig.~\ref{fig:spectra_dw_3d} are nearly degenerate. This arises because, relative to the $x$ and $y$~directions, the symmetry along the $z$~axis is only \emph{slightly} broken by the presence of the DW.
%\nob{However, contrarily to the Q1D setup,
%some of the states
%shown in Fig.~\ref{fig:spectra_dw_3d}
%for the 3D trap
%some of the previous states 
%\nog{have energies}
%are nearly
%} \nog{almost}
%degenerate,
%\CB{XXXXXX}
%\nog{with others with excitations in~$x$ and~$y$ directions} 
%\nog{as}\CG{since compared to the $x$ and~$y$ directions} the symmetry in the~$z$ direction is
%only \emph{slightly} broken due to the presence of the DW \CG{in this direction}.}
%\nog{In this case}
For example,
while the state 
$\varphi^{\rm RM, LBS}_{{\rm DW}, a_g}  \psi^{\rm CM}_{{\rm DW}, a_g}(0, 0, 2)$ (magenta)
is 
%\nog{unique}
non-degenerate, the states
$\varphi^{\rm RM, LBS}_{{\rm DW}, a_g}  \psi^{\rm CM}_{{\rm DW}, a_g}(2, 0, 0)$ and
$\varphi^{\rm RM, LBS}_{{\rm DW}, a_g}  \psi^{\rm CM}_{{\rm DW}, a_g}(0, 2, 0)$ (green)
are degenerate, 
and
%\nog{being}
their energies are very similar to those of the state
$\varphi^{\rm RM, LBS}_{{\rm DW}, a_g}  \psi^{\rm CM}_{{\rm DW}, a_g}(0, 0, 4)$ (also green).
Similarly,
the state
$\varphi^{\rm RM, LBS}_{{\rm DW}, a_g}  \psi^{\rm CM}_{{\rm DW}, a_g}(0, 0, 6)$ (yellow),
has an energy that is only slightly smaller than those of the degenerate states
\linebreak
$\varphi^{\rm RM, LBS}_{{\rm DW}, a_g}  \psi^{\rm CM}_{{\rm DW}, a_g}(2, 0, 2)$
and 
$\varphi^{\rm RM, LBS}_{{\rm DW}, a_g}  \psi^{\rm CM}_{{\rm DW}, a_g}(0, 2, 2)$
(brown).

%\CG{[AS: I think this part is rather confusing. Maybe one should first discuss the three shown 
%avoided crossings. (Question: is there also a true crossing as discussed by Simon in the?) 
%Then discussion of the width compared to the anisotropic case.]}

As 
%\nog{inferred from}
can be seen in the enlarged view of the avoided crossings,
the values of their widths are
several orders of magnitude smaller than those
for the Q1D setup
(note the different scales of the insets in Fig.~\ref{fig:spectra_dw_q1d}
compared to those in the inset in Fig.~\ref{fig:spectra_dw_3d}).
These avoided crossings involve 
the first trap state (Mott insulator)
$\varphi^{\rm RM, TS1}_{{\rm DW}, a_g}  \psi^{\rm CM}_{{\rm DW}, a_g}(0, 0, 0)$ 
and the molecular states
$\varphi^{\rm RM, LBS}_{{\rm DW}, a_g}  \psi^{\rm CM}_{{\rm DW}, a_g}(0, 0, 2)$ (magenta),
$\varphi^{\rm RM, LBS}_{{\rm DW}, a_g}  \psi^{\rm CM}_{{\rm DW}, a_g}(2, 0, 0)$,
and
$\varphi^{\rm RM, LBS}_{{\rm DW}, a_g}  \psi^{\rm CM}_{{\rm DW}, a_g}(0, 2, 0)$
(green),
$\varphi^{\rm RM, LBS}_{{\rm DW}, a_g}  \psi^{\rm CM}_{{\rm DW}, a_g}(0, 0, 4)$
(green),
and
$\varphi^{\rm RM, LBS}_{{\rm DW}, a_g}  \psi^{\rm CM}_{{\rm DW}, a_g}(0, 0, 6)$ (yellow).
We believe that the dramatic difference
in the widths of the avoided crossings
is responsible of the absence of dispersive features in the 3D 
experiment, 
in agreement with previous related predictions by
Bolda \emph{et al.}~\cite{cold:bold05}
and
Schneider \emph{et al.}~\cite{cold:schn09}.
Thus, experimentally observing these ICIRs would require a very
fine tuning of the interparticle
interactions at a much slower 
%\nog{fashion} \CG{
pace to follow the corresponding adiabatic
curves, 
according to the Landau-Zener theory~\cite{cold:land32, cold:zen32}.

%
%--------------------------------------------------------------------------------------------
% Figure: 3D SPECTRA
%------------------------------------------------------------.--------------------------------
%--------------------------------------------------------------------------------------

%
%==================================================================
\subsection{Validity
%\nog{Validation}
and limits of
%\nog{the}  
single-well models
%\nog{descriptions}
} \label{sec:results_sw}
%==================================================================

\begin{figure}[hbtp!]
%\includegraphics[width=\columnwidth]{figs/aniso_harm-swAg-B1u_4300-4300-1315_colored}
% Spectrum with the double well
\includegraphics[width=0.95\columnwidth]{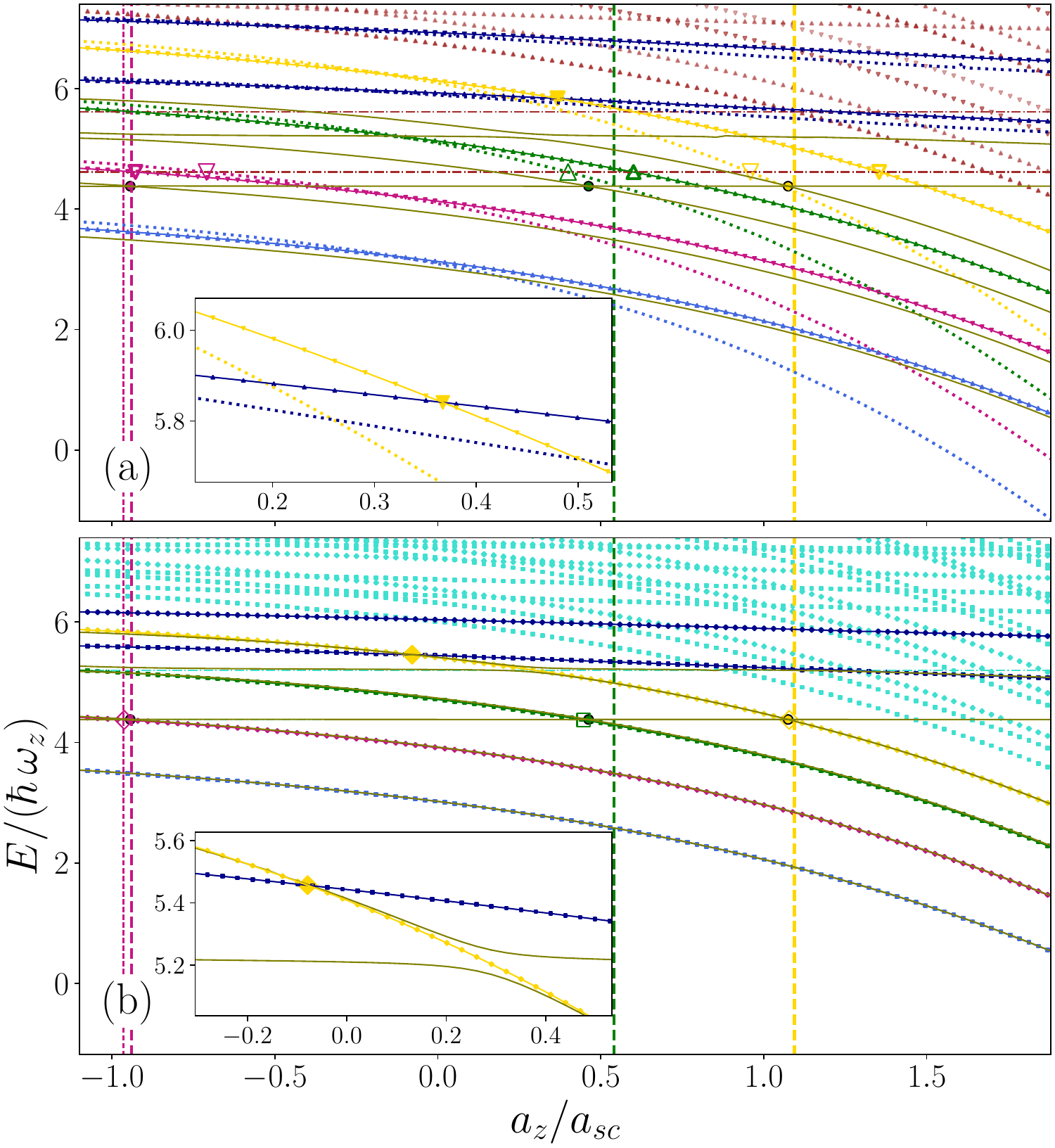}
\caption{
Same as Fig.~\ref{fig:spectra_dw_q1d}
%\nog{and~\ref{fig:spectra_dw_3d}}, 
but for the eigenstates of the $A_g$ and $B_{1u}$ irreducible representations 
%\nog{associated with two atoms
%confined in an anisotropic optical lattice
%with trap depths
%$V_x = V_y = 35 \,E_R$,
%and~$V_z = 10.7 \, E_R$ 
%, if described by}
and an isotropic single-well model using
the Taylor expansion given by Eq.~\eqref{eq:Vsw}
up to order (a) 2 (harmonic approximation),
with some of the results rendered by the 
regularized $\delta$ pseudopotential superimposed (dotted lines),
and (b) 6 (sextic approximation). 
The solid olive lines show some of the
energies of the
$A_g$ double-well states of Fig.~\ref{fig:spectra_dw_q1d}.
}
\label{fig:spectra_harm-sw_q1d}
\end{figure}

\begin{figure}[hbtp!]
\includegraphics[width=0.95\columnwidth]{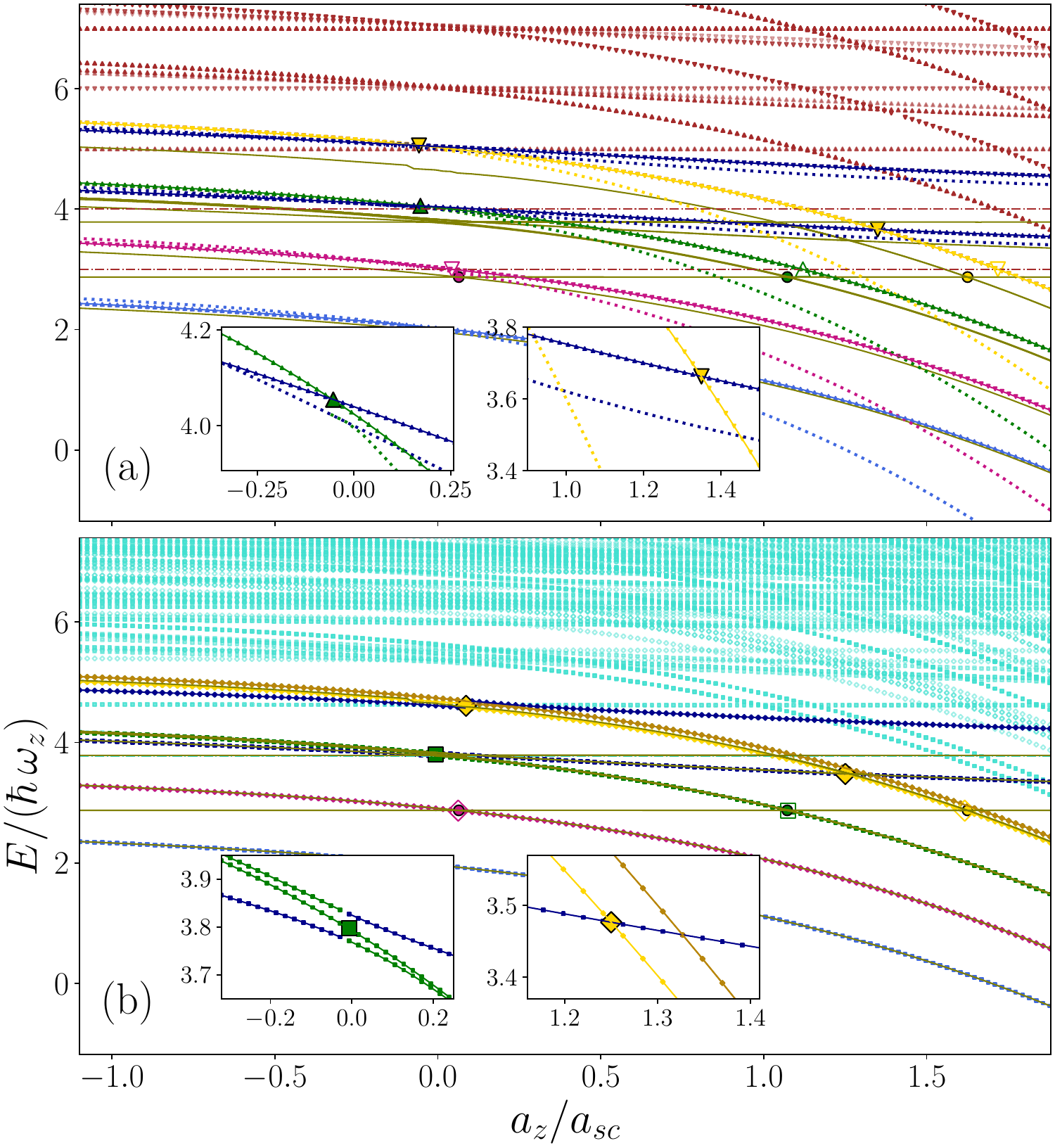}
\caption{Same as Fig.~\ref{fig:spectra_harm-sw_q1d}, but for
an isotropic trap with
$V_x = V_y = V_z = 35 \, E_R$.
}
\label{fig:spectra_harm-sw_3d}
\end{figure}

In this section, we discuss 
the appropriateness of a SW description of the system 
to predict the dispersive features 
reported in the experiment in Ref.~\cite{cold:lee23}.
For this purpose,
%by providing ICIRs associated with avoided crossing at 
%adequate positions.
%\nog{we show} 
the energy spectrum is shown for
%\nog{associated with}
the anisotropic Q1D setups  
%\nog{for}
SWM2 (harmonic) and SWM6 (sextic) 
%models 
%for the harmonic (SW2) and sextic  (SW6) SW models 
%are shown
in Figs.~\ref{fig:spectra_harm-sw_q1d}(a)
and~\ref{fig:spectra_harm-sw_q1d}(b),
respectively.
%\nog{Here}
%\nob{For comparison,
%the DWM eigenenergies of interest
%[cf. Figs.~\ref{fig:spectra_dw_q1d}
%and~\ref{fig:spectra_dw_3d}]
%have been also superimposed.}
%previously discussed.
The corresponding results for the
isotropic 3D trap
%\fr{Notice that the DW in the $z$ direction
%makes the model anisotropic}
%\CG{[AS: the caption of Fig.\,7 says 
%"isotropic". I think the confusion comes from the fact that 
%the focus of Figs. 6 and 7 is on the single-well results 
%(shown here the first time), but also the corresponding 
%DW results are superimposed for comparison, and this latter 
%information is given just in the previous sentence. Maybe one 
%should first introduce the two Figs., i.e., anisotropic and 
%isotropic cases described within SWMs, and then write that for 
%comparison also the corresponding DWM results are shown?]}
are presented in
Figs.~\ref{fig:spectra_harm-sw_3d}(a)
and~\ref{fig:spectra_harm-sw_3d}(b).
For comparison,
the DWM eigenenergies of interest
[cf.\ the colored curves in Figs.~\ref{fig:spectra_dw_q1d}
and~\ref{fig:spectra_dw_3d}]
have been also superimposed.
For the sake of clarity,
we show only those SW eigenstates
that are totally symmetric with respect to all Cartesian planes,
i.\,e., they belong to the $A_g$ IR,
and those that are
symmetric with respect to the
$xz$ and $yz$ Cartesian planes,
and antisymmetric with respect 
to the~$xy$ plane, i.\,e., 
they belong to the $B_{1u}$ IR~\cite{gen:rotman95}.

The discussion, 
and thus this section, 
is divided in five parts.
First, %\nog{briefly discuss}
the reproduction of a Mott insulator state 
with SW models is briefly discussed
in Sec.~\ref{sec:results_sw_trap_states}.
Second, Sec.~\ref{sec:results_sw_crossings}
is devoted to the SW predictions,
as well as their implications on the 
symmetry conservation.
Third, Sec.~\ref{sec:results_sw_multiwell}
details how the SW limitations can be
overcome to deal with multiwell phenomena.
Fourth, Sec.~\ref{sec:results_sw_delta}
summarizes the performance of a
regularized $\delta$
%\nog{regularized}
interatomic potential.
Fifth, Sec.~\ref{sec:results_sw_positions}
contains the main findings of this 
work by comparing the ICIR positions
of all considered models, and their relation to the
experimental features.

%
%--------------------------------------------------------------------------------------------
% Figure : SW, DW (Ix=Iy=17200mW/cm^2, Iz=5258mW/cm^2)
%--------------------------------------------------------------------------------------------
%--------------------------------------------------------------------------------------------

%
%--------------------------------------------------------------------------------------------
% Figure : SW, DW (Ix=Iy=17200mW/cm^2, Iz=5258mW/cm^2)
%--------------------------------------------------------------------------------------------
%--------------------------------------------------------------------------------------------

%==================================================================
\subsubsection{Presence and absence of trap states}
%under single-well approximations} 
\label{sec:results_sw_trap_states}
%==================================================================
%Trap states
%First,
%\nog{Notice that t}
The energy branch of the first trap state
obtained within the DWM
%DW model
(black lines in Figs.~\ref{fig:spectra_dw_q1d}
and~\ref{fig:spectra_dw_3d}) and
that plays the role of a Mott insulator in an infinite 
optical lattice (without short-range interatomic interactions)
is absent in both SW models,
%\nog{.}  
%\nog{A}\CG{,
as can be seen in
%\nog{inferred from} \CG{seen in} 
Figs.~\ref{fig:spectra_harm-sw_q1d}
and~\ref{fig:spectra_harm-sw_3d}.
Instead, 
the \emph{first} trap state of the SW, i.\,e., 
$\varphi^{\rm RM, TS1}_{{\rm SW}, a_g}  \psi^{\rm CM}_{{\rm SW}, a_g}(0, 0, 0)$
(lower dark blue curve) 
%\fr{sounds more natural}
%\nob{(dark blue, lower curve),}
which belongs to the $A_g$ IR~\cite{cold:busc98},
is similar to the \emph{second}
$A_g$ trap state
%\nog{of the DW} 
$\varphi^{\rm RM, TS2}_{{\rm DW}, a_g} \psi^{\rm CM}_{{\rm DW}, a_g}(0, 0, 0)$
of the DWM.
In fact, the probability densities of both states are
also very similar, as
%\nog{can be concluded}
is seen from the comparison of the 
%\nog{sections of} 
cuts through the probability densities shown in
Figs.~\ref{fig:wf_mott}(a') and~\ref{fig:wf_mott}(a).
The next trap state in ascending order of energy is
$\varphi^{\rm RM, TS1}_{{\rm SW}, a_g}  \psi^{\rm CM}_{{\rm SW}, b_{1u}}(0, 0, 1)$ (upper dark blue curve) and belongs to the $B_{1u}$ IR.
The corresponding probability density shown in
Fig.~\ref{fig:wf_mott}(b')
has a node at the origin.
The probability density for both SWM6 states
is mostly concentrated along the anti-diagonal~$z_2=-z_1$
(in the case of the $B_{1u}$ trap state slightly off-set due to the
existence of a node along that line),
which translates into a large value of the RM
coordinate and, thus, to a large separation.
However, the effects of the short-range interaction are
noticeable and responsible for the probability density to be
elongated along the diagonal~$z_2=z_1$.

%==================================================================
\subsubsection{Diabatic and adiabatic crossings,
and symmetry considerations}
%under single-well approximations} 
\label{sec:results_sw_crossings}
%==================================================================
%\newpage
%Second,
%\nog{Notice that}
Noteworthy, in the Q1D setup
shown in Fig.~\ref{fig:spectra_harm-sw_q1d}
the only crossing between
%with
the lowest-lying trap state
$\varphi^{\rm RM, TS1}_{{\rm DW}, a_g}  \psi^{\rm CM}_{{\rm DW}, a_g}(0, 0, 0)$
%(bottom dark blue line)
and the colored molecular branches
involves the molecular state
$\varphi^{\rm RM, LBS}_{{\rm SW}, a_g}  \psi^{\rm CM}_{{\rm SW}, b_{1u}}(0, 0, 3)$ (yellow). Remind, however, that the 
DWM exhibits more crossings [cf.\, Fig.~\ref{fig:spectra_dw_q1d}].
As can be seen in the insets of Figs.~\ref{fig:spectra_harm-sw_q1d}
and \ref{fig:spectra_harm-sw_3d},
the mentioned crossing is a \emph{true (diabatic)}
crossing due to the different symmetries 
of the interacting states
($B_{1u}$ for the molecular state
and $A_g$ for the trap state).
This diabatic crossing also appears in the 3D case
(see right insets of Figs.~\ref{fig:spectra_harm-sw_3d}(a)
and~\ref{fig:spectra_harm-sw_3d}(b)).
Contrarily,
the \emph{corresponding} crossing for the DWM is avoided
[cf.\ the olive lines in the inset of 
Fig.~\ref{fig:spectra_harm-sw_q1d}(b)].
Recall that under such an isotropic trapping potential,
the molecular state
$\varphi^{\rm RM, LBS}_{{\rm SW}, a_g}  \psi^{\rm CM}_{{\rm SW}, b_{1u}}(0, 0, 3)$
is degenerate with the states
$\varphi^{\rm RM, LBS}_{{\rm SW}, a_g}  \psi^{\rm CM}_{{\rm SW}, b_{2u}}(0, 3, 0)$
and
$\varphi^{\rm RM, LBS}_{{\rm SW}, a_g}  \psi^{\rm CM}_{{\rm SW}, b_{3u}}(3, 0, 0)$.
In all cases,
%In other words,
the mentioned crossings cannot be associated with an 
ICIR, since the crossings are true ones, i.\,e., not avoided crossings.
However, in the 3D case more crossings 
with the lowest-lying trap state
$\varphi^{\rm RM, TS1}_{{\rm SW}, a_g}  \psi^{\rm CM}_{{\rm SW}, a_g}(0, 0, 0)$
are present, as can be seen in Fig.~\ref{fig:spectra_harm-sw_3d}.
For example, the molecular states
$\varphi^{\rm RM, LBS}_{{\rm SW}, a_g}  \psi^{\rm CM}_{{\rm SW}, a_g}(0, 0, 2)$
(green) and
$\varphi^{\rm RM, LBS}_{{\rm SW}, a_g}  \psi^{\rm CM}_{{\rm SW}, b_{1u}}(0, 0, 3)$
(yellow)
cross the trap states
$\varphi^{\rm RM, TS1}_{{\rm SW}, a_g}  \psi^{\rm CM}_{{\rm SW}, a_g}(0, 0, 0)$
(lower dark blue curve),
%\nob{(dark blue, lower curve),}
and
$\varphi^{\rm RM, TS1}_{{\rm SW}, a_g}  \psi^{\rm CM}_{{\rm SW}, b_{1u}}(0, 0, 1)$
(top dark blue), respectively.
Furthermore, the left insets of Fig.~\ref{fig:spectra_harm-sw_3d}
show a dramatic difference between the previous crossings
for SWM2 and SWM6:
%for the SW2 and SW6 models:
while within the harmonic approximation
%[see Fig.~\ref{fig:spectra_harm-sw_3d}(a/b)]
the crossings are
%\nog{\emph{real}}
\emph{true} ones,
the anharmonicities present in the SWM6
%SW6 model
force the crossings between states with the same symmetry to be avoided,
as they concern states with the same symmetry~\cite{cold:vonNeum29}.
Notice in Fig.~\ref{fig:spectra_harm-sw_3d}(b) that the energy of
the (yellow) state
$\varphi^{\rm RM, LBS}_{{\rm SW6}, a_g}  \psi^{\rm CM}_{{\rm SW6}, b_{1u}}(0, 0, 3)$\footnote{Note that while the state $\varphi^{\mathrm{RM, LBS}}_{\mathrm{SW6}, a_g} \psi^{\mathrm{CM}}_{\mathrm{SW6}, b_{1u}}(0, 0, 3)$ is non-degenerate within the $b_{1u}$ irreducible representation, it is degenerate in energy with the states $\varphi^{\mathrm{RM, LBS}}_{\mathrm{SW6}, a_g} \psi^{\mathrm{CM}}_{\mathrm{SW6}, b_{2u}}(0, 3, 0)$ and $\varphi^{\mathrm{RM, LBS}}_{\mathrm{SW6}, a_g} \psi^{\mathrm{CM}}_{\mathrm{SW6}, b_{3u}}(3, 0, 0)$, which are antisymmetric with respect to the $xz$ and $yz$ Cartesian planes, respectively.
},
which is 
%\nog{unique}
non-degenerate and very similar to that of the state
$\varphi^{\rm RM, LBS}_{{\rm DW}, a_g}  \psi^{\rm CM}_{{\rm DW}, a_g}(0, 0, 6)$,
is only slightly smaller than that of the degenerate
states (brown) 
$\varphi^{\rm RM, LBS}_{{\rm SW6}, a_g}  \psi^{\rm CM}_{{\rm SW6}, b_{1u}}(2, 0, 1)$
and
$\varphi^{\rm RM, LBS}_{{\rm SW6}, a_g}  \psi^{\rm CM}_{{\rm SW6}, b_{1u}}(0, 2, 1)$,
%\nog{which are degenerate} 
and also very similar to 
%\nog{those} 
the energies of the states
$\varphi^{\rm RM, LBS}_{{\rm DW}, a_g}  \psi^{\rm CM}_{{\rm DW}, a_g}(2, 0, 2)$
and
$\varphi^{\rm RM, LBS}_{{\rm DW}, a_g}  \psi^{\rm CM}_{{\rm DW}, a_g}(0, 2, 2)$. 
These results stem from the structural similarities among the constituent orbitals. On the one hand, the RM orbitals $\varphi^{\mathrm{RM, LBS}}_{\mathrm{SW6}, a_g}$ and $\varphi^{\mathrm{RM, LBS}}_{\mathrm{DW}, a_g}$ are closely matched,
both displaying wild oscillations near the origin.
On the other hand, their CM orbitals
also share similar spatial profiles:
each features two excitations in either the $x$ or $y$ direction and
a single excitation per well along $z$.
Specifically, the SWM6 states
$\psi^{\mathrm{CM}}_{\mathrm{SW6}, b_{1u}}(2, 0, 1)$ and $\psi^{\mathrm{CM}}_{\mathrm{SW6}, b_{1u}}(0, 2, 1)$
contain a single $z$ excitation overall within the (single) well,
whereas the DWM states
$\psi^{\mathrm{CM}}_{\mathrm{DW}, a_g}(2, 0, 2)$ and $\psi^{\mathrm{CM}}_{\mathrm{DW}, a_g}(0, 2, 2)$ possess two $z$ excitations,
where one is localized in each well.

%==================================================================
\subsubsection{Multiwell effects
described within single-well
approximations}
%under single-well approximations} 
\label{sec:results_sw_multiwell}
%==================================================================

%\newpage
%Third,
%Still,
The SW models can be used to analyze multiwell effects~\cite{cold:lee23},
if the zero\CG{-}energy point related to the Mott insulator is
included \emph{ad hoc} in the energy spectrum.
This energy can be approximated by the
ground-state energy of the non-interacting case
$(a_z/a_{\rm sc} \rightarrow \infty)$.
Then, the 
%\nog{non-existing}
(artificially introduced) 
crossings between the
energy of this manually added state and
%\nog{those} 
the ones of the molecular branches, which have been marked 
with empty symbols in Figs.~\ref{fig:spectra_harm-sw_q1d}
and~\ref{fig:spectra_harm-sw_3d},
can be effectively used to
%\nog{identify}
predict
ICIRs involving the Mott insulator state.
The reason is twofold.
On the one hand,
the zero-crossing energy
for the SW, 
especially for the 
SWM6,
%SW6 model,
lies very close to the energy of the state
$\varphi^{\rm RM, TS1}_{{\rm DW}, a_g}  \psi^{\rm CM}_{{\rm DW}, a_g}(0, 0, 0)$.
On the other hand,
the energies of the SW molecular states, 
especially in the SWM6 case,
are very similar to the ones of the DWM molecular states. 
Concretely, the energies of the states
$\varphi^{\rm RM, LBS}_{{\rm DW}, a_g}  \psi^{\rm CM}_{{\rm DW}, a_g}(0, 0, n_{\rm DW}^{\rm CM})$ 
are very similar to those of the states
$\varphi^{\rm RM, LBS}_{{\rm SW6}, a_g}  \psi^{\rm CM}_{{\rm SW6}, a_g}(0, 0, n_{\rm SW}^{\rm CM})$,
and
$\varphi^{\rm RM, LBS}_{{\rm SW}, a_g}  \psi^{\rm CM}_{{\rm SW}, b_{1u}}(0, 0, n_{\rm SW}^{\rm CM})$,
depending on whether~$n_{\rm SW}^{\rm CM} = n_{\rm DW}^{\rm CM}/2$
is even or odd, respectively~\footnote{%
%-----------------------------------------------------------------------------
% FOOTNOTE
%-----------------------------------------------------------------------------
The lowest-lying eigenenergies of the 
DWM appear 
as nearly degenerate pairs with the same 
RM part,
$\varphi^{\rm RM,LBS}_{{\rm DW},a_g}$,
but with an additional excitation in 
the CM coordinate:
%which also changes their symmetry
$\psi^{\rm CM}_{{\rm DW},a_g}(0,0,n_{\rm DW}^{\rm CM})$ 
(wavefunctions of the $a_g$ IR, i.\,e.,
totally symmetric) and
$\psi^{\rm CM}_{{\rm DW},b_{1u}}(0,0, n_{\rm DW}^{\rm CM}+1)$
(wavefunctions
of the $b_{1u}$ IR, i.\,e., 
symmetric with respect to the Cartesian planes~$xz$ and~$yz$,
and antisymmetric with respect to the plane~$xy$), 
where $n_{\rm DW}^{\rm CM}$ is an even integer. 
The different symmetries of the molecular states
$\varphi^{\rm RM,LBS}_{{\rm DW},a_g}
\psi^{\rm CM}_{{\rm DW},b_{1u}}(0,0, n_{\rm DW}^{\rm CM}+1)$ 
and the lowest-lying trap state
$\varphi^{\rm RM,TS1}_{{\rm DW},a_g}
\psi^{\rm CM}_{{\rm DW},a_g}(0,0,0)$
that plays the role of
%\nog{a}
the lowest-lying Mott-insulator state
prohibits their interaction.
%-----------------------------------------------------------------------------
}.
The agreement between the SW6 and DW eigenenergies is also
manifested in the remarkable similarities between their 
molecular states shown in Fig.~\ref{fig:wf_rmlbs-CM0-3}
within one of the wells,
especially for those that are less affected by the 
anharmonicities in the optical potential due to
their smaller number of CM excitations. 
In the 3D case, more states are degenerate due to 
the higher symmetry.

\begin{figure}[hbtp!]
\includegraphics[width=0.75\columnwidth]{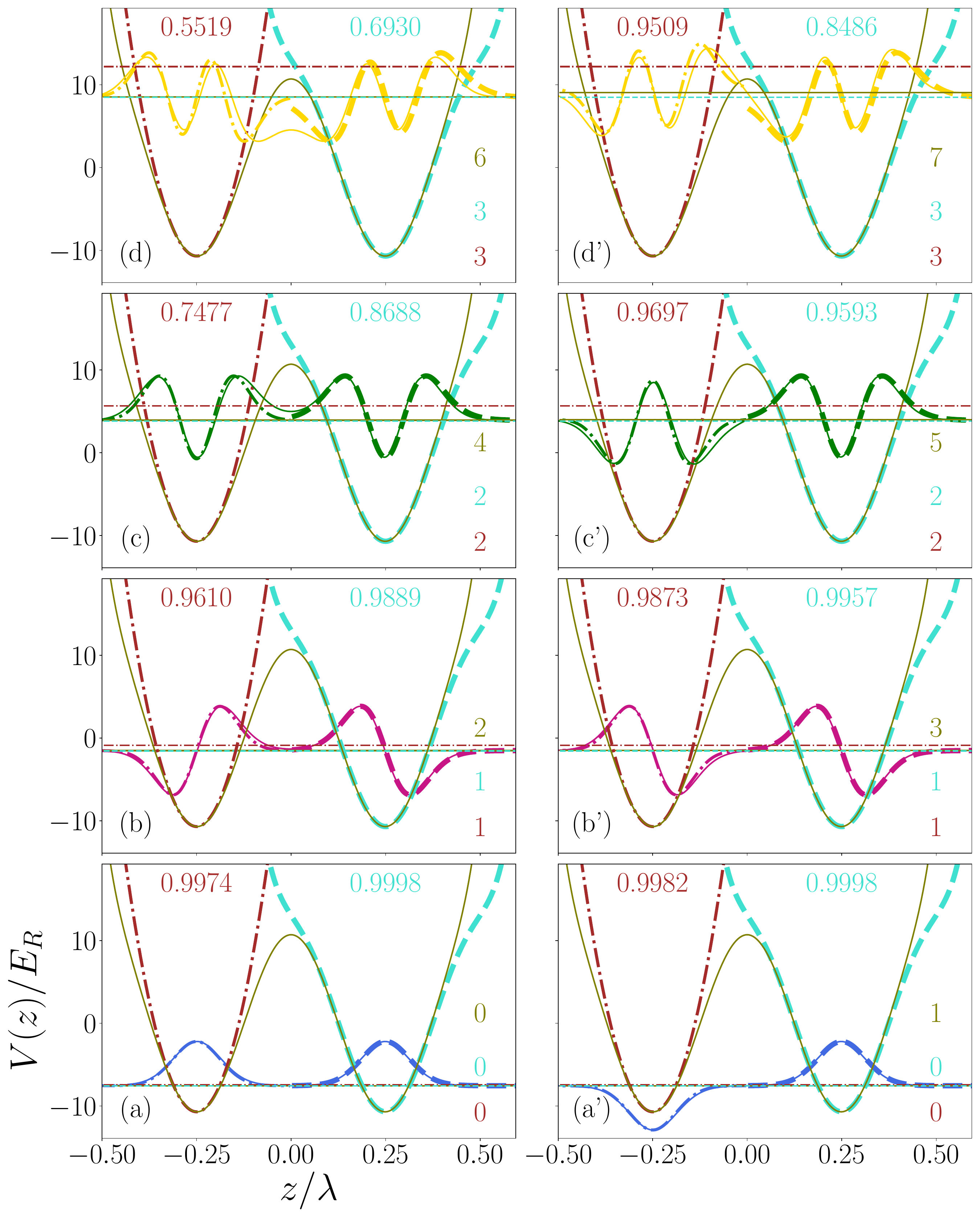}
\caption{Cuts along the $z$ direction of the lowest-lying
center-of-mass eigenstates of the Hamiltonian~\eqref{eq:HCM}
for a quasi-one-dimensional optical lattice with 
depth~$V_z = 10.7 \, E_R$.
The lattice is modeled using three different potentials:
a single-well harmonic approximation~\eqref{eq:Vharm} (dash-dotted brown, left wells), 
a single-well sextic potential~\eqref{eq:Vsw} (dashed cyan, right wells),
and a double-well potential~\eqref{eq:Vdw} (thin solid olive).
The eigenstate profiles are plotted using line styles matching those
of their respective model potentials and colored according to the
total eigenstates shown in Fig.~\ref{fig:wf_rmlbs-CM0-3}.
The number of excitations for each eigenstate is indicated in the lower right corner of each panel, and
the overlaps of the single-well states with the double-well states on the top. The eigenenergies are shown (with the same 
color and line style) as horizontal lines. The (arbitrary 
phases) of the SW eigenstates are chosen to match the 
corresponding DW eigenstates.
}
\label{fig:pot_aniso}
\end{figure}

To unravel the similarities 
and the differences between the SW (SW2 and SW6) and DW 
models in more detail, Fig.~\ref{fig:pot_aniso} shows cuts 
along the~$z$ direction of the
lowest-lying CM eigenstates that form those total eigenstates
(see Table~\ref{tab:states_vs_colors})
that are responsible for the previously discussed ICIRs.
Here, both the symmetric (left column) and the antisymmetric 
(right column) CM eigenstates for the DWM are plotted.  
As can be seen, the overlaps confirm that the DWM eigenstates agree more closely with the SWM6 eigenstates than with those of SWM2.
This better agreement is also observed for the corresponding eigenenergies.
Notice, also,
that this agreement is particularly remarkable
between the SWM6 and the DWM lowest-lying eigenstates.
These results indicate the importance of the anharmonicities
in the optical trap, and the consequent limitations of the 
harmonic approximation. Similarly, note
that the SWM eigenstates show a better agreement 
with the antisymmetric than with the
symmetric DWM eigenstates. This fact can be explained due to the existence of a central node
in the antisymmetric eigenstates of the DWM,
which nullifies the probability density of the wave function,  
mimicking better the situation of a SW where 
the density is by construction decaying exponentially outside 
the well.

Finally, notice that the DW states shown 
in the left column of Fig.~\ref{fig:pot_aniso}
can (approximately) be obtained by the addition
and subtraction of the SW states (right column),  
a common feature of
DW potentials, such as the one experienced by
the electron of the hydrogen molecular ion, 
there explaining the occurrence of 
bonding and antibonding orbitals. 
This concept of 
superposition of single-well 
eigenfunctions connects the present results also
with
%\nog{those previously discussed} 
the lattice-induced-resonance model
by Stecher \emph{et al.} in Ref.~\cite{cold:stec11},
where symmetric and antisymmetric
Wannier wavefunctions are considered
to explain the formation of molecular dimers
by the association of two atoms confined
in neighboring sites of optical lattices.

%\newpage
Remarkably, although totally symmetric ($A_g$)
two-particle states confined in a 3D 
optical lattice can be constructed with 
combinations of RM and CM orbitals with other symmetries, 
e.\,g., with RM and CM orbitals
where both belong to
%\nog{associated with}
the
$b_{1g}$ IR,
only the fully symmetric RM and CM orbitals %\nog{yield} 
comprise a molecular bound state that can potentially couple to the trap states. Other symmetries yield either 
deeply bound states that do not enable 
this coupling or trap states where 
the two atoms are far apart.
Thus, for the occurrence of ICIRs
involving the first-trap state
that plays the role
of a Mott insulator state in an infinite
optical lattice, only the $a_g$ CM symmetric states 
(see cuts along the $z$ axis in the left column of 
Fig.~\ref{fig:pot_aniso}) are relevant.  
Contrarily, the two-particle wave function
for the CM antisymmetric states 
(right column of Fig.~\ref{fig:pot_aniso})
do not present any avoided crossing
with the mentioned first-trap state
due to their strong energetic difference in
the RM part.

%
%--------------------------------------------------------------------------------------------
% Figure: ANISOTROPIC POTENTIALS
%--------------------------------------------------------------------------------------------
%--------------------------------------------------------------------------------------
%

%==================================================================
\subsubsection{
$\delta$ 
contact pseudopotential}
%\nob{potential}}
%under single-well approximations} 
\label{sec:results_sw_delta}
%==================================================================

%pseudopotential
%Fourth,
Another very popular approximation in describing ultracold atoms is 
to model the interatomic interaction by a (regularized) $\delta$ 
contact potential. This was also used in the modeling of the experiment in \cite{cold:lee23}.
As can be seen in Figs.~\ref{fig:spectra_harm-sw_q1d}(a)
and~\ref{fig:spectra_harm-sw_3d}(a),
within the attractive regime ($a_{\rm sc}<0$)
the energy levels obtained with the $\delta$ potential
are very similar to the ones of the \emph{ab-initio} results.
%\nog{,
%which have been superimposed with the same colors 
%vas were used in the %Figs.~\ref{fig:spectra_harm-sw_q1d}(a)
%and~\ref{fig:spectra_harm-sw_3d}(a)}%
Conversely, the finite-size effects of the interaction
become noticeable within the repulsive regime ($a_{\rm sc}>0$),
especially for small interactions, i.\,e.,
small absolute values of $a_{\rm sc}$, where the pseudopotential 
pronouncedly underestimates the \emph{ab-initio} results,  
see also \cite{cold:wulf26}. 
Still, in the Q1D setup, the crossing points of the first-trap state of the DWM with the molecular states 
calculated using the $\delta$ pseudopotential (and a harmonic 
trap) are in surprisingly good agreement with 
both the DWM and the experimental
results for~$n_{\rm SW2}^{\rm CM} = 2$ (green) and~3 (yellow). 
This is due to error cancellation,
as the smaller molecular energy for the  pseudopotential
compensates the larger energy of the harmonic trap with
respect to SWM6 and DWM. 
(Note the increasing energy difference 
in Fig.~\ref{fig:pot_aniso}
as the number of CM excitations rises.)
As a consequence, the 
%\nog{provided}
positions of the avoided crossings, and thus
%\nog{then}
of the corresponding ICIRs,
are accidentally 
agreeing better with experiment in the 
case of 
%\nog{more adequate for} 
the pseudopotential than
%\nog{for}
the \emph{ab-initio} 
results.
%\nog{curves}.
The positions of the crossings 
are also substantially different for the 3D trap, but here 
no experimental results exist for comparison.
Notice, however, that the results rendered by the 
pseudopotential
would be worse than those of the
\emph{ab-initio} interaction for other models,
such as SWM6 or DWM, which provide, nonetheless, a more 
appropriate description of the optical lattice.

%==================================================================
\subsubsection{Positions of the avoided crossing
for the double- and single-well models}
\begin{figure}[hbtp!]
\includegraphics[width = 0.75 \columnwidth]{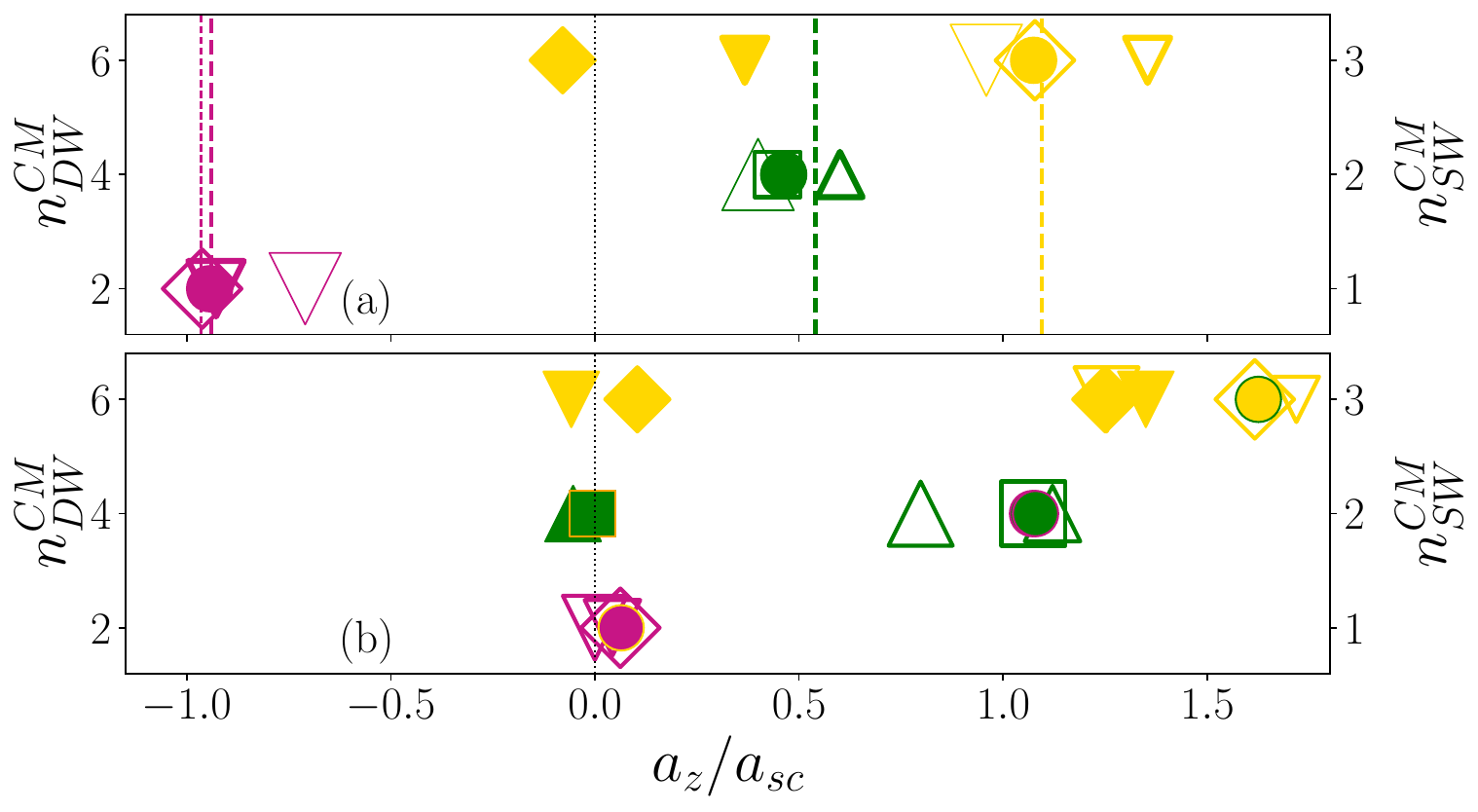}
\caption{Positions of the avoided crossings 
for the optical trap with a depth 
of~$V_x = V_y = 35 \, E_R$ and 
(a)~$V_z = 10.7 \, E_R$
or
%\nog{and}
(b)~$V_z = 35 \, E_R$,
with their the number of center-of-mass excitations
for the double-well of Figs.~\ref{fig:spectra_dw_q1d}
and~\ref{fig:spectra_dw_3d} (left axis),
shown as circles,
and the single-well models (right axis)  
of Figs.~\ref{fig:spectra_harm-sw_q1d} and~\ref{fig:spectra_harm-sw_3d}, shown as diamonds and 
squares (sextic potential) and
triangles (harmonic approximation), respectively.
The filled symbols correspond to \emph{true} avoided crossings 
between the molecular states 
$\varphi^{\rm RM, LBS}_{{\rm DW/SW}, a_g}  \psi^{\rm CM}_{{\rm DW/SW}, a_g}(0, 0, n^{\rm CM}_{\rm DW/SW})$,
with~$n^{\rm CM}_{\rm SW} = n^{\rm CM}_{\rm DW}/2 = 1, 2, 3$,
and the first trap state
$\varphi^{\rm RM, TS1}_{{\rm DW/SW}, a_g}  \psi^{\rm CM}_{{\rm DW/SW}, a_g}(0, 0, 0)$ 
as they appear in the energy spectrum. 
The empty symbols correspond to the values of 
$a_z/a_{\rm sc}$ where the energy of the corresponding 
molecular state agrees with the one of non-interacting 
atoms (which emulates the perfect Mott-insulator ground state.)
The interatomic potential
is described using \emph{ab-initio} curves,
except for
the (thin) triangles, which 
correspond to
the $\delta$ contact pseudopotential.
}
\label{fig:icir_positions}
\end{figure}

%\nog{descriptions}}
%under single-well approximations} 
\label{sec:results_sw_positions}
%==================================================================
%
Figure~\ref{fig:icir_positions} 
shows the positions of the crossings observed
in the SWM and DWM spectra.
This figure summarizes the main results of this work.
Recall that while the SW models (SW2: filled yellow triangle,
SW6: filled yellow diamond) present only one true
crossing involving the first trap state
$\varphi^{\rm RM, TS1}_{{\rm SW}, a_g}  \psi^{\rm CM}_{{\rm SW}, a_g}(0, 0, 0)$ and
$\varphi^{\rm RM, LBS}_{{\rm SW}, a_g}  \psi^{\rm CM}_{{\rm SW}, b_{1u}}(0, 0, 3)$,
the DWM shows three avoided crossing (filled circles)
in agreement with the experimental features (dashed lines). 
However, the three crossings of the zero-energy point of the 
SWM2 (empty triangles) show a moderate 
agreement with the experimental and with the DWM results.
This agreement is substantially better for the SWM6
(empty diamonds and squares), a result that 
highlights the importance of anharmonicities in the optical lattice.
Note, also, that due to the already discussed cancellation of errors
the $\delta$ pseudopotential (thin triangles)
yields more accurate results 
than the \emph{ab-initio} interacting potential
(thick triangles)
for 
$\varphi^{\rm RM, LBS}_{{\rm SW}, a_g}  \psi^{\rm CM}_{{\rm SW}, a_g}(0, 0, 2)$ (green)
and 
$\varphi^{\rm RM, LBS}_{{\rm SW}, a_g}  \psi^{\rm CM}_{{\rm SW}, b_{1u}}(0, 0, 3)$ (yellow),
but a less accurate result for
$\varphi^{\rm RM, LBS}_{{\rm SW}, a_g}  \psi^{\rm CM}_{{\rm SW}, b_{1u}}(0, 0, 1)$ (magenta).
Of course, only the avoided crossings (absent in the SW2 approximation) reveal also 
the coupling strengths and thus the widths of the ICIRs.

The comparison between the positions of the avoided crossings
(filled symbols),
i.\,e., of the ICIRs, for the Q1D and 3D setups shown in
Figs.~\ref{fig:icir_positions}(a) and ~\ref{fig:icir_positions}(b), 
respectively, also renders interesting results.
On the one hand, the ICIR that involves 
the molecular state
\linebreak
$\varphi^{\rm RM, LBS}_{{\rm DW}, a_g}  \psi^{\rm CM}_{{\rm DW}, a_g}(0, 0, 2)$, whose energy is very close to that of
$\varphi^{\rm RM, LBS}_{{\rm SW6}, a_g}  \psi^{\rm CM}_{{\rm SW6}, b_{1u}}(0, 0, 1)$, and the trap state
%\linebreak
$\varphi^{\rm RM, TS1}_{{\rm DW}, a_g}  \psi^{\rm CM}_{{\rm DW}, a_g}(0, 0, 0)$
moves from the attractive branch of the spectrum
\linebreak
($a_{\rm sc} = -\,2651$\,{\au}, $a_z / a_{\rm sc} = -0.944$)
%-0.943788
to a position \emph{close} to the unitary limit
within the repulsive branch of the spectrum (magenta symbols). 
The corresponding scattering length is one order of magnitude larger than in the Q1D case.
On the other hand, due to the \emph{stronger} confinement in the 3D case,
the other two ICIRs associated with the molecular states
$\varphi^{\rm RM, LBS}_{{\rm DW}, a_g}  \psi^{\rm CM}_{{\rm DW}, a_g}(0, 0, 4)$ (green circle)
and
$\varphi^{\rm RM, LBS}_{{\rm DW}, a_g}  \psi^{\rm CM}_{{\rm DW}, a_g}(0, 0, 6)$ (yellow circle),
whose energies are similar to those of the states
$\varphi^{\rm RM, LBS}_{{\rm SW6}, a_g}  \psi^{\rm CM}_{{\rm SW6}, a_g}(0, 0, 2)$
and
$\varphi^{\rm RM, LBS}_{{\rm SW6}, a_g}  \psi^{\rm CM}_{{\rm SW6}, b_{1u}}(0, 0, 3)$,
respectively, appear for substantially 
larger interactions
in the Q1D setup than in the 3D case,
as the values of the corresponding scattering lengths
are much 
larger
($a_{\rm sc} = 5411$\,{\au} vs $a_{\rm sc} = 1724$\,{\au},
i.\,e., $a_z/a_{\rm sc} = 0.462$ vs $1.079$,
for
$\varphi^{\rm RM, LBS}_{{\rm DW}, a_g}  \psi^{\rm CM}_{{\rm DW}, a_g}(0, 0, 4)$,
and
$a_{\rm sc} = 2328$\,{\au} vs $a_{\rm sc} = 1145$\,{\au},
i.\,e., $a_z/a_{\rm sc} = 1.075$ vs $1.625$,
for
$\varphi^{\rm RM, LBS}_{{\rm DW}, a_g}  \psi^{\rm CM}_{{\rm DW}, a_g}(0, 0, 6)$).

%\nob{($a_{\rm sc} = 2840$\,{\au} vs $a_{\rm sc} = 1237$\,{\au}
%for
%$\varphi^{\rm RM, LBS}_{{\rm DW}, a_g}  \psi^{\rm CM}_{{\rm %DW}, a_g}(0, 0, 4)$,
%and
%$a_{\rm sc} = 1224$\,{\au} vs $a_{\rm sc} = 815$\,{\au}
%for
%$\varphi^{\rm RM, LBS}_{{\rm DW}, a_g}  \psi^{\rm CM}_{{\rm %DW}, a_g}(0, 0, 6)$).}

%
%--------------------------------------------------------------------------------------------
% Figure : SW, DW (Ix=Iy=Iz=17200mW/cm^2)
%--------------------------------------------------------------------------------------------
%--------------------------------------------------------------------------------------

%\newpage . \newpage
%==================================================================
\subsection{Alternative pictures for explaining 
 the experimentally observed resonance features}
\label{sec:results_discuss}
%==================================================================
%
As already mentioned, the authors of Ref.~\cite{cold:lee23} 
relate their experimental results to lattice-induced 
resonances~\cite{cold:stec11} and emphasize 
that they
cannot be confinement-induced resonances 
(see the introduction and, especially,  footnote [25] in Ref.~\cite{cold:lee23}).
The main argument for this 
claim is based on symmetry, namely that the experimental features are,
within the adopted model, ascribed to 
odd-parity excitations that should be symmetry forbidden. 
The symmetry breaking in lattice-induced resonances 
is explained by interaction-driven tunneling. Since 
the present work claims ICIRs to be responsible for 
the resonances observed in Ref.~\cite{cold:lee23} 
and also in view of earlier confusion regarding 
confinement-induced resonances, some comments 
appear to be necessary. 

First of all, the name \emph{inelastic} confinement-induced 
resonances (ICIR) was introduced in 
Ref.~\cite{cold:sala12} in order to distinguish 
those confinement-induced resonances that are based 
on CM-RM coupling between a bound molecular state 
and an unbound trap state from the ones that involve 
only the RM and may be understood as a divergence 
in the effective interaction (scattering length) due 
to dimensional confinement to
Q1D~\cite{cold:olsh98,cold:berg03} 
or Q2D~\cite{cold:petr00}
geometry
which were originally termed 
\emph{confinement-induced resonances} and later 
renamed into \emph{elastic} confinement-induced 
resonances \cite{cold:sala12}. The most striking 
difference between the two types of CIRs is that 
only in the case of ICIRs molecule formation 
or breaking is possible, since the binding energy 
can be transferred in between RM and CM degrees of 
freedom, as was predicted in Ref.~\cite{cold:sala12} 
and experimentally confirmed in Ref.~\cite{cold:sala13} 
where overlapping elastic and inelastic CIRs were observed 
and distinguished. With the given definition, CIR 
include elastic and inelastic CIRs, while inelastic 
ICIRs include among others anharmonic CIRs as well 
as lattice-induced resonances as special cases.

Second, as discussed in  
Sec.~\ref{sec:results_sw_crossings},
the claimed symmetry violation 
of an ICIR-based explanation
in Ref.~\cite{cold:lee23} is due to a limitation 
of the adopted model where the crossing of the 
eigenstates of a single-well potential with the 
non-interacting state is adopted. As is demonstrated, 
already a DWM for describing two neighbour sites of a lattice does not encounter symmetry issues.
Noteworthy, in the here
adopted model 
%\nog{here} \CG{[AS: I think the original 
%(restored) position of "here" is correct.]}
the resonant features occur as 
\emph{avoided} crossings in the \emph{ab-initio} energy 
spectrum, confirming the coupling between the involved 
molecular bound state and the trap state. If the 
coupling would vanish due to symmetry, a true instead 
of an avoided crossing would occur. The physical 
picture behind is that, quantum mechanically, there 
is the same probability for the molecule to form 
either in the left or in the right well. In fact, 
the obtained state is an (equal) superposition 
of the molecule being left or right. (It may be noted 
that this is also directly visible in the 2nd quantized 
formulation of the Hamiltonian given in 
Ref.~\cite{cold:stec11}.) 

Third, it may be noted that the common solid-state 
physics formulation and language (hopping, tunneling) 
as is also adopted in Ref.~\cite{cold:lee23} infer   
a time-dependent picture, even for the description 
of stationary states. This is in contrast to the 
stationary approach adopted in this work 
for the analysis of the energy spectra. This leads to different 
\emph{pictures} describing, however, the same physical 
process. In the experiment the atoms are initially 
prepared in a singly-occupied
Mott insulator state. 
Then, the magnetic field is varied resulting in a 
time-dependent variation of the two-particle 
interaction. The formation of a molecule requires 
the two atoms to be sufficiently close to each other, 
in the present case to be on the same lattice site. 
In one picture, this requires (active) tunneling of 
one of the atoms from its original site to the 
neighbor site, denoted as interaction-induced 
tunneling in Ref.~\cite{cold:lee23}. (In 
2nd quantization language, the atom is 
annihilated on one site and created on its 
neighbor site.)
In the complementary picture, the 
molecule formation is possible
even in a singly-occupied Mott state 
if there is a
non-zero probability for the atom to occupy the neighbor site.
Mathematically, 
the delocalization (imperfect localization of the 
ground-state Wannier function) and the hopping 
(tunneling) matrix elements are directly related, 
thus leading to the same molecule-formation 
probability. It may be noted that active tunneling 
from one site to the other would occur, e.\,g., if 
the lattice would be time-dependently tilted or 
the barrier between the wells would otherwise be 
varied~\cite{cold:mein14}. 

Fourth, the present findings fully support the claim 
in Ref.~\cite{cold:lee23} that in the experiment 
(also to the authors' knowledge for the first time) 
the molecule formation of atoms 
%\nog{originally located at} 
involving
%\CG{[AS: In view of the fact that we emphasize earlier 
%that the Wannier functions are delocalized over more than 
%one well, my original formulation is not really sensible.]}
different lattice sites has been observed. Furthermore, 
the present work shows that a quantitative modeling of 
the experimental results is possible already adopting 
a double-well model.
(Remind that
it was checked that 
a triple-well model did effectively not change the 
results.) Again, this finding is not surprising in 
view of the fact that the experiment started with 
a Mott-insulator state. This implies relatively localized 
Wannier functions and narrow bands. The hopping amplitudes 
are thus small relative to the on-site terms. As a 
consequence, the tight-binding approximation considering 
only next-neighbor interactions is a good approximation. 
Consequently, a double-well model should also be 
a reasonable approximation. While it may be very 
interesting to observe also "true" lattice resonances 
where atoms delocalized over the whole lattice may be 
turned into molecules, the corresponding coupling 
is expected to be very small \cite{cold:bold05}.

%\newpage . \newpage
%==================================================================
% SUMMARY AND OUTLOOK
\section{Summary and outlook} \label{sec:summary}
%==================================================================

In this work,
we provide a comprehensive alternative explanation for the novel dispersive features
experimentally observed by Lee \emph{et al.}\ in Ref.~\cite{cold:lee23}
in ultracold atoms trapped in an optical lattice. 
%\nog{After analyzing}
Based on the analysis of various 
optical-lattice models,
from idealized harmonic to complex multiwell configurations,
we demonstrate that the experimental results are accurately explained by a trapping potential modeled with anharmonic 
single-wells in two directions and a double-well in the third.
Specifically, the resonance features observed experimentally 
under quasi-one-dimensional trapping show very good agreement  
with the avoided crossings in  
the calculated  two-particle spectrum.
These avoided crossings involve molecular states 
(combining the relative-motion least-bound state
with lowest-lying center-of-mass excitations)
and a unique first-trap state of the double-well.
This first-trap state naturally represents a perfect Mott insulator with single-site occupancy,
a crucial aspect not captured by single-well descriptions of the optical lattice.
Furthermore, Lee \emph{et al.}\ attributed 
this phenomenon to lattice-induced resonances facilitated by tunneling.
Our framework reinterprets the resonant features
as inelastic confinement-induced resonances, which manifest as avoided crossings in the
two-particle energy spectrum.
We demonstrate that these avoided crossings,
even when involving states delocalized across multiple wells,
%\nog{and thus implying tunneling},
arise from the same fundamental mechanism:
the coupling between the relative-motion and center-of-mass
%\nog{coordinates}
degrees of freedom due to the trapping potential's anharmonicities. 
Therefore, we provide a unified view of both 
anharmonicity-induced resonances~\cite{cold:kest10, cold:kest10b},
lattice-induced~\cite{cold:stec11},
and
inelastic confinement-induced resonances~\cite{cold:sala12},
as we conclude that they share the same underlying physical origin. 
The ICIRs that were defined as resonances associated with 
avoided crossings between a molecular state and a trap state 
caused by the coupling between center-of-mass and relative 
motion may be seen as the general phenomenon, including 
anharmonicity-induced as well as lattice-induced resonances 
as special cases.

Furthermore, our model
%\nog{explains} 
provides a possible 
explanation for the absence of similar resonance features
under three-dimensional trapping 
for the discussed
%\nog{conceived}
experiment, 
attributing it to the significantly narrower width of the corresponding avoided crossings in 
the two-particle spectrum~\cite{cold:bold05, cold:schn09}.
This suggests that observing multiwell
inelastic confinement-induced resonances 
under three-dimensional confinement\footnote{Note that ICIRs
under 3D confinement involving two atoms at the same site
have recently been reported in Ref.~\cite{cold:cape23}.
}
would require much finer tuning of interparticle interactions. 

While the experimental Mott-insulator state is not fully captured by single-well descriptions,
we also investigated the feasibility of such models in predicting 
inelastic confinement-induced resonances positions.
We found that the double-well spectrum's avoided crossings align remarkably well
with the crossings of the single-well molecular states
(formed by the least-bound state and the lowest-lying 
center-of-mass one
%\nog{coordinates})
with the energy of the non-interacting state,
%\nog{energy},
which effectively resembles %\nog{represents}
the zero-point energy associated with a perfect Mott insulator state.
Our analysis further underscores the critical role of anharmonicities in the trapping potential,
as evidenced by the improved accuracy of 
sixth-order
%\nog{sextic-}
Taylor series expansions over harmonic approximations.
Notably, the double-well description resolves an apparent symmetry violation present in single-well models,
providing a more complete and accurate representation of the system. 

In summary, this work significantly advances the understanding of multiwell effects in optical lattices,
which certainly opens new routes to control and manipulate novel correlated many-body states.

%}

%==================================================================
% ACKNOWLEDGEMENTS
\section*{Acknowledgments}
\label{sec:thanks}
%==================================================================

We thank E.~Lee, H.~Lin, and W.~Ketterle for useful discussions and for providing the experimental data~\cite{cold:lee23} prior to publication. We also extend our gratitude to Manuele Landini and Christoph Nägerl for insightful discussions. 
The authors acknowledge computing resources at the Magerit Supercomputer
of the Universidad Polit\'{e}cnica de Madrid. 

Generative AI tools were used to improve the clarity and readability of the manuscript.
All content generated using these tools was critically reviewed, verified,
and edited by the authors, who accept full responsibility for the final version of the article.

%==================================================================
% CrediT
%==================================================================
%\paragraph{Author contributions}

F.R.\ conceived the study and
performed the numerical simulations.
F.R.\ and A.S.\ analyzed the numerical results and wrote the manuscript.

%==================================================================
% FUNDING
%==================================================================
\paragraph{Funding information}
This work has been supported by the Grant PID2021-122711NB-C21
funded by MCIN/AEI/10.13039/501100011033. 
AS gratefully acknowledges 
support by Berlin Quantum.

%==================================================================
% BIBLIOGRAPHY
%==================================================================
%\bibliographystyle{apsrev4-1}    
%\bibliographystyle{aipsamp}% Produces the bibliography via BibTeX.
%\bibliographystyle{apsrev4-1}
%\bibliographystyle{unsrt}
%\nocite{*}
\bibliographystyle{SciPost_bibstyle}
\bibliography{journals,cold,cold_new,gen_new}  %gen

\end{document}